\pdfoutput=1
\documentclass[11pt,a4paper]{article}
\usepackage{jinstpub}

\usepackage[T1]{fontenc}
\usepackage{microtype}
\usepackage{siunitx}
\usepackage{booktabs}
\usepackage{multirow}
\usepackage{array}
\usepackage{subcaption}
\usepackage{xspace}
\usepackage{listings}
\usepackage{xcolor}
\usepackage{tabularx}
\graphicspath{{figures_log/}{figures/}{figures_qsimfix/}{figures_2p5GeV/}{figures_5xulp/}{figures_muon/}{figures_proton/}{figures_calorimetry/}}

\newcommand{\Simphony}{Simphony\xspace}
\newcommand{\Opticks}{Opticks\xspace}
\newcommand{\Geant}{Geant4\xspace}
\newcommand{\code}[1]{\texttt{#1}}
\newcommand{\chisqndf}{\ensuremath{\chi^2/\mathrm{ndf}}\xspace}

\title{GPU optical photon Monte Carlo for noble liquid detectors:
validation against Geant4 in a large liquid argon TPC benchmark}

\author[a,1]{Gabor Galgoczi,\note{Corresponding author.}}
\author[a]{Xuyang Ning,}
\author[a]{Dmitri Smirnov,}
\author[a]{Brett Viren}
\author[a]{and Chao Zhang}

\affiliation[a]{Brookhaven National Laboratory,\
Upton, NY 11973, U.S.A.}

\emailAdd{ggalgoczi1@bnl.gov}

\abstract{%
Optical photon Monte Carlo simulation is a computational bottleneck for noble liquid Time Projection Chambers. Design studies require repeated, geometry dependent simulations of timing, wavelength shifting, and optical response, while reconstruction and particle identification workflows need labeled optical datasets. We present \Simphony, a GPU optical simulation tool, formerly EIC-Opticks, built on \Opticks with CUDA and NVIDIA OptiX. \Simphony implements a GPU version of the \Geant{} \code{G4OpWLS} wavelength-shifting model and returns Monte Carlo truth for detected hits with low per-photon overhead. We validate \Simphony against \Geant{} 11.3.2 in a simplified \SI{14.7}{kt} liquid argon Time Projection Chamber benchmark with a two-stage wavelength-shifting shell and idealized photon counting detector. For three paired \SI{2.5}{GeV} electron simulations, each producing about \SI{61}{M} optical photons, the integrated detected-photon ratio agrees with \Geant{} at the subpercent level. The detected arrival time and wavelength spectra give \chisqndf values of 0.98 and 1.08. Contained muon and near-Cerenkov-threshold proton samples give $R_N=1.0017\pm0.0008$ and $R_N=1.0005\pm0.0014$, confirming agreement for distinct source topologies. On an NVIDIA RTX 4090, a stacked launch of four \SI{2.5}{GeV} electron events transports 243 M optical photons in $3.03\pm0.06$ s, giving $80.2\pm1.6$ M photons s$^{-1}$. Relative to a single-thread \Geant{} reference and including GPU overheads and host-device transfers, the optical transport speedup is $1053\pm55$; the end-to-end wall time acceleration is $89\pm5$. These results show that \Simphony can make explicit optical photon Monte Carlo practical for detector development studies and for generating machine learning optical response datasets.}


\begin{document}
\maketitle
\flushbottom

\pagebreak

\section{Introduction}
\label{sec:introduction}

\subsection{Optical simulation needs in large noble liquid TPCs}
\label{sec:intro-motivation}

Optical photon simulation is an important but computationally expensive component of large noble liquid Time Projection Chamber (TPC) studies. In liquid argon, energy deposition produces both ionization charge and scintillation light. The charge signal provides the primary imaging and calorimetric observable in present LArTPC neutrino experiments, while the light signal is commonly used for triggering, event timing, and charge--light matching. Future detector studies place more emphasis on detailed predictions of the light response, including photon arrival times, wavelength shifting, optical boundary effects, and the geometry dependent response of photon detectors.
% References for these?  "Future" -> "Recent"?

The computational challenge comes from the photon multiplicity. A GeV-scale interaction in liquid argon can produce on the order of ten million optical photons. In large detectors, these photons may propagate over tens of meters of distance, undergo Rayleigh scattering, encounter many optical boundary interactions, and pass through one or more wavelength-shifting stages before detection. The expense of implementing optical transport with CPU-based \Geant{}~\cite{geant4_2003, geant4_2016} multiplies for detector designers that must vary the geometry and material properties to optimize detector concepts.

The need for high throughput optical simulation is growing due to increasing emphasis of the role that scintillation light can play. Low-energy electron studies in The Liquid Argon In A Testbeam (LArIAT) experiment have shown that including scintillation information can improve
calorimetric reconstruction relative to methods using charge only ~\cite{foreman2020lariatLightCalorimetry}.  Recent GeV-scale studies indicate that light calorimetry in LArTPCs can be self-compensating,
because recombination transfers part of the heavily ionising hadronic response from charge into light~\cite{ning2025lartpc}. At MeV energies, enhanced photon detection is relevant for solar neutrino measurements, supernova neutrino burst
detection, and other low energy neutrino applications
~\cite{capozzi2019duneSolar,shi2025mevArgon,zhu2019mevBackgrounds,
dune2021supernovaBurst,dune2025supernovaPointing,moller2018dsnb}.
These physics goals motivate detector concepts with increased and more uniform light collection, including upgraded photon detectors considered for future, large LArTPC  modules~\cite{dune2024phaseII,marinho2025apex}, where optimizing these detector concepts demands large scale optical simulation studies.

\subsection{Fast optical workflows and GPU transport}
\label{sec:intro-fast-optical}

Several fast optical workflows have been developed to avoid repeated full optical photon Monte Carlo CPU transport. Lookup tables, semi-analytic response models and neural network surrogates can replace detailed optical tracking with a precomputed or learned detector response
~\cite{sbnd2024light,garciagamez2021transport,mu2022photon}. These methods are often well suited to production simulation once the detector geometry, optical properties, and photon detector layout are fixed. In LArSoft, for example, fast optical modules store optical channel visibilities with respect to optical voxels, reducing the need to propagate every optical photon with \Geant{}~\cite{snider2017larsoft,larsoftPDFastSimPAR}.

The trade-off is that response models must be regenerated, retrained, or revalidated when the detector geometry, optical properties, or nuisance parameter priors change. Lookup tables can also introduce voxel size effects, memory pressure, and reduced timing or truth information. These limitations are particularly relevant for machine learning workflows, where large labeled samples are needed not only for training, but also for systematic variations and validation.

Table~\ref{tab:optical-simulation-strategies} summarizes the main workflow trade-offs among common optical simulation strategies. The comparison is qualitative rather than a ranking: the best choice depends on detector size, optical channel count, timing requirements, output content, hardware, and the amount of revalidation required after geometry or material
changes. Full CPU optical Monte Carlo remains the most direct reference calculation, but its repeated per-event cost is high for large noble liquid detectors. Lookup tables, response maps, and neural network surrogates shift much of this cost into an offline generation or training stage, they can be very fast once the detector model is fixed, but must be regenerated, retrained, or revalidated when the geometry, optical properties, photon detector layout, or systematic assumptions change
~\cite{garciagamez2021transport,mu2022photon,lei2022siren}. For example, an ICARUS scale optical library with about \(2\times10^{6}\) sampling points was reported to require about one week to generate~\cite{lei2022siren}.

GPU optical Monte Carlo is complementary to these approaches. It keeps explicit photon level transport, timing, wavelength shifting, and hit truth information, while mapping the large optical photon multiplicity onto GPU parallelism~\cite{blyth2019opticks,wenzel2024cats}. It is therefore useful when the response model itself is changing, or when first principles samples are needed to build, validate, or update photon libraries and surrogate models.

\begin{table}[t]
  \centering
  \small
  \setlength{\tabcolsep}{4pt}
  \renewcommand{\arraystretch}{1.18}
  \caption{Workflow level comparison of optical simulation strategies. Entries indicate typical trade-offs relevant to detector design and
  production studies.}
  \label{tab:optical-simulation-strategies}
  \begin{tabularx}{\linewidth}{@{}
    >{\raggedright\arraybackslash}p{0.21\linewidth}
    >{\raggedright\arraybackslash}X
    >{\raggedright\arraybackslash}X
    >{\raggedright\arraybackslash}p{0.21\linewidth}
    @{}}
    \toprule
    \textbf{Strategy}
      & \textbf{Main strength}
      & \textbf{Update or resource cost}
      & \textbf{Typical use} \\
    \midrule

    Full CPU optical MC
      & Direct photon level reference with full timing, wavelength, and truth
        information available.
      & High repeated per-event cost for large photon counts.
      & Reference calculations and small high fidelity samples. \\

    Lookup table or response map
      & Fast production response after the library has been generated.
      & Regeneration or revalidation after geometry, material, or detector
        layout changes, storage can be large.
      & Fixed geometry production simulation. \\

    ML surrogate
      & Fast inference, can be useful when a differentiable response model is
        needed.
      & Requires training data and validation; significant detector changes
        may require retraining.
      & Fast response emulation and ML-integrated workflows. \\

   GPU optical MC
      & Explicit photon transport with timing, wavelength shifting, and
        hit truth retained at lower repeated cost than CPU optical
        MC for high photon count events.
      & Requires GPU geometry  integration 
      & Detector scans, lookup table validation, and surrogate training
        sample generation. \\

    \bottomrule
  \end{tabularx}
\end{table}

This GPU Monte Carlo strategy was pioneered for particle physics optical
simulation by the \Opticks framework, which combines \Geant{} event
simulation with NVIDIA OptiX ray tracing and CUDA based optical physics
~\cite{blyth2019opticks}. The simplified benchmark geometry used in this
paper is a validation choice rather than a restriction of the underlying
geometry model: \Opticks implements ray intersections for analytic CSG
primitive shapes and represents more complex detector solids through Boolean
CSG combinations~\cite{blyth2019opticks}.  The same geometry machinery has
been exercised in JUNO scale detector geometries with large numbers of
repeated PMT assemblies and structural volumes~\cite{blyth2024opticks}.
The approach has also been demonstrated in \Geant{}--\Opticks integration
workflows, Ring Imaging Cherenkov detectors, and low photon yield optical
transport applications
~\cite{wenzel2024cats,li2023lhcbOpticks,galgoczi2025eicopticks}.

\subsection{\Simphony and scope of this work}
\label{sec:intro-simphony}

\Simphony is the renamed and extended continuation of EIC-Opticks~\cite{galgoczi2025eicopticks}. The rename reflects the broader scope of the package beyond the Electron Ion Collider, including noble liquid and LArTPC optical transport applications. EIC-Opticks adapted
the \Opticks workflow to low- and moderate-yield detector workloads by aggregating multiple \Geant{} events into a single GPU launch, and validated the method in the ePIC pfRICH geometry. \Simphony keeps the
\Opticks/OptiX/CUDA foundation and extends it for large noble liquid optical
simulations.

A central addition in this work is a GPU wavelength shifting implementation matched to the \Geant{} \code{G4OpWLS} model. The implementation uses the same \Geant{} material property interface, including
\code{WLSCOMPONENT}, \code{WLSABSLENGTH}, and \code{WLSTIMECONSTANT}, and
performs WLS absorption, re-emission, timing, direction, and polarization
sampling on the GPU. The implementation follows
the \Geant{} WLS model and is validated against \Geant{}.

\begin{figure}[h!]
  \centering
  \includegraphics[width=0.78\linewidth]{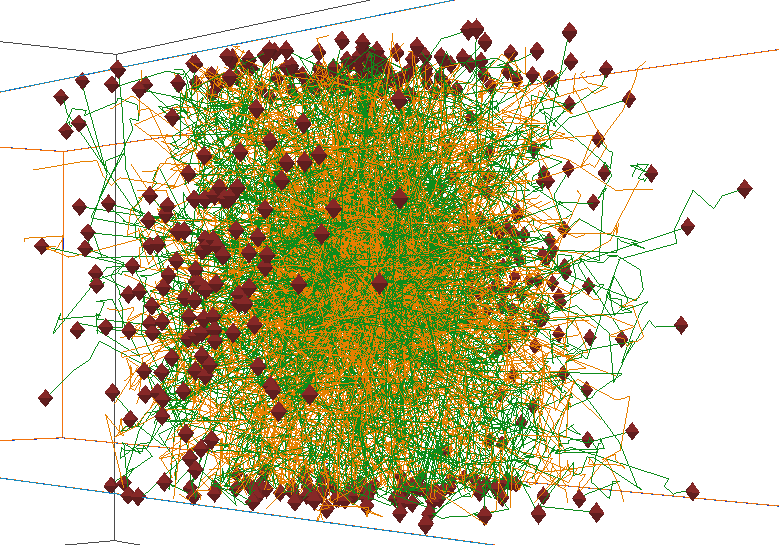}
  \caption{Photons simulated by Simphony on GPU generated by a \SI{0.5}{MeV} electron event in a 14.7 kt liquid argon detector. Only photons (green) that were detected and photons (orange) that touched the wavelength-shifting layers but were not detected are shown in the event. Red markers show the detection points. The GeV-scale benchmark events studied below launch approximately \(6\times10^7\) optical photons, about \(6\times10^4\) times more than are drawn in this visualization.}
  \label{fig:opticalphot}
\end{figure}

We validate \Simphony in a simplified large liquid argon TPC benchmark with a \SI{60}{m} $\times$ \SI{13.5}{m} $\times$ \SI{13}{m} active volume, corresponding to \SI{14.7}{kt} of liquid argon. The argon volume is surrounded by a two stage wavelength-shifting shell and an idealized detector with 100\% photon detection efficiency. This geometry preserves a
large detector optical transport scale and a realistic two stage WLS cascade, but deliberately avoids detector specific photon detector details. It should therefore be interpreted as a controlled transport benchmark, not as a performance model of any particular experiment. Figure~\ref{fig:opticalphot} shows an example GPU optical photon event display in the benchmark geometry. 

The paper compares \Simphony with \Geant{} 11.3.2 using identical initial photon distributions created by electron, muon, and proton sources.  We report
hit number agreement, arrival time and hit wavelength spectrum
comparisons, spatial hit map comparisons, GPU throughput, end-to-end timing, and
memory use on an NVIDIA RTX 4090. We also demonstrate the practical use of the accelerated transport with a four hour optical calorimetry parameter
scan. The validation establishes agreement with \Geant{} for the simplified benchmark and WLS model studied here, further work is needed for realistic photon detector response, experiment specific production integration, and validation against measured data.

\section{Simulation model and benchmark configuration}
\label{sec:simulation-model}

\subsection{Benchmark geometry}
\label{sec:geometry}

The benchmark geometry is a rectangular liquid argon volume with dimensions
\SI{60}{m} $\times$ \SI{13.5}{m} $\times$ \SI{13}{m}. For a liquid argon density of \SI{1.396}{g\,cm^{-3}}, this corresponds to an active mass of \SI{14.7}{kt}. The argon volume is enclosed by three nested optical layers:
a \SI{200}{\micro m} para-terphenyl (pTP) wavelength-shifting layer, a \SI{6}{mm} TPB-doped acrylic wavelength-shifting layer, and a
\SI{1}{mm} outer photon detector shell.  The outer shell is assigned \SI{100}{\percent} detection efficiency over the full optical band and is therefore an idealized photon counting boundary rather than a model of a
specific photon detector.

The geometry is designed as a controlled optical transport benchmark. It retains a representative \SI{15}{kt}-scale argon volume and a two stage wavelength-shifting chain, while deliberately omitting detector specific features such as segmented modules, cathode or membrane placement, dichroic or filter surfaces, SiPM spectral response, a realistic photon detection efficiency.

\begin{figure}[h!]
  \centering
  \includegraphics[width=0.9\linewidth]{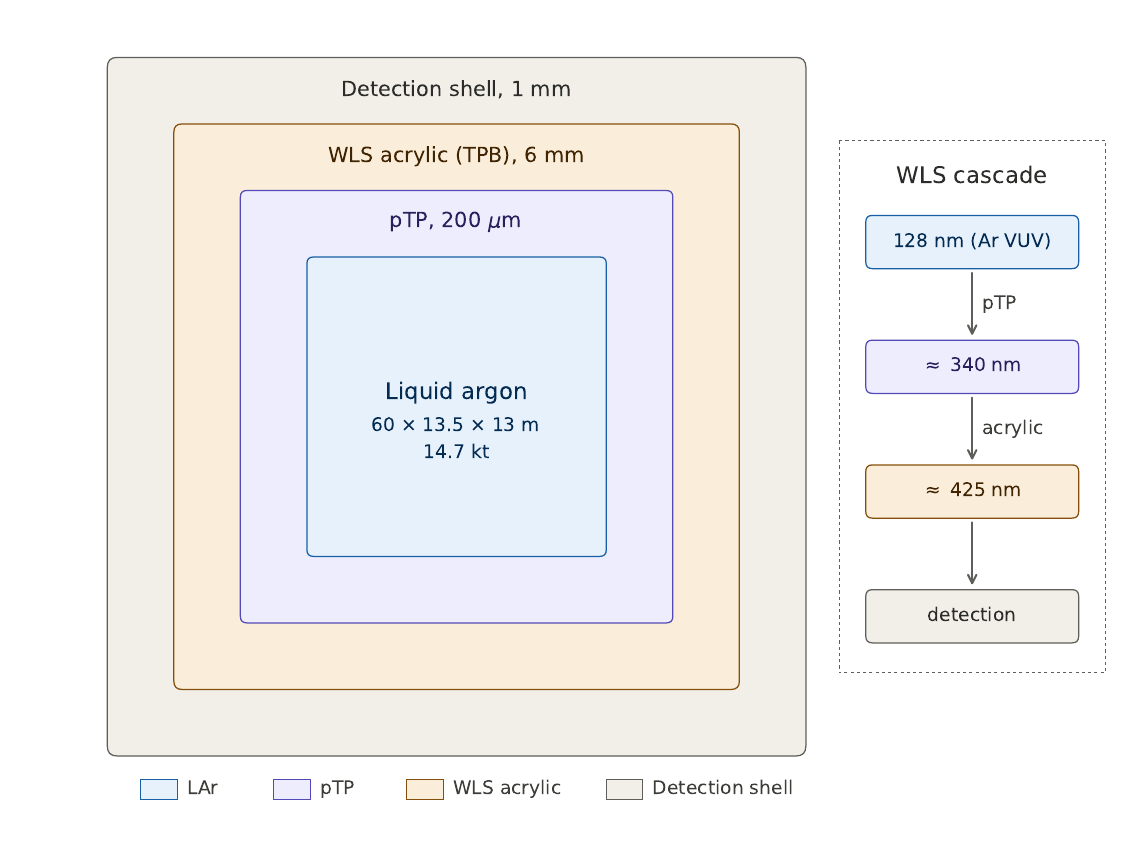}
  \caption{Simplified \SI{14.7}{kt} liquid argon TPC benchmark geometry. The left panels show the LAr active volume and the nested WLS/detection layers. The right panel illustrates the two stage wavelength-shifting path, from \SI{128}{nm} to approximately \SI{340}{nm} in pTP and from approximately \SI{340}{nm} to approximately \SI{425}{nm} in the
  wavelength-shifting acrylic.}
  \label{fig:geometry}
\end{figure}

Table~\ref{tab:geometry} summarizes the simplified benchmark geometry. The simplified benchmark geometry is shown in figure~\ref{fig:geometry}.

\begin{table}[t]
  \centering
  \small
  \setlength{\tabcolsep}{4pt}
  \renewcommand{\arraystretch}{1.15}
  \caption{Simplified benchmark geometry. The photon detector is an idealized sensitive boundary.}
  \label{tab:geometry}
  \begin{tabular}{@{}
    >{\raggedright\arraybackslash}p{0.16\linewidth}
    >{\raggedright\arraybackslash}p{0.20\linewidth}
    >{\raggedright\arraybackslash}p{0.29\linewidth}
    >{\raggedright\arraybackslash}p{0.22\linewidth}
    @{}}
    \toprule
    Region & Thickness or size & Main optical role & Comment \\
    \midrule
    Liquid argon
      & \SI{60}{m} $\times$ \SI{13.5}{m} $\times$ \SI{13}{m}
      & Scintillation, Rayleigh scattering, absorption
      & \SI{14.7}{kt} at \SI{1.396}{g\,cm^{-3}} \\
    pTP
      & \SI{200}{\micro m}
      & WLS, VUV to near-UV
      & First WLS stage \\
    WLS acrylic
      & \SI{6}{mm}
      & WLS, near-UV to blue
      & Second WLS stage \\
    Detection shell
      & \SI{1}{mm}
      & Photon counting
      & Idealized boundary; \code{EFFICIENCY}=1.0 \\
    \bottomrule
  \end{tabular}
\end{table}

\subsection{Optical model}
\label{sec:optical-model}

The optical processes included in the benchmark are scintillation and Cerenkov photon generation, bulk absorption, Rayleigh scattering, optical boundary interactions, and wavelength shifting. The CPU and GPU configurations use the same material property tables and optical surface settings wherever the corresponding process is implemented in both stacks. The boundary processes use the same UNIFIED model optical surface parameters
in the two simulations.

The liquid argon scintillation model contains fast and slow components with relative amplitudes 0.75 and 0.25 and decay constants of \SI{7}{ns} and \SI{1400}{ns}, respectively.  The WLS materials are defined through
\Geant{} material property tables specifying the WLS absorption length, emission spectrum, and decay time. In \Simphony, the WLS process is a GPU implementation matched to the \Geant{} \code{G4OpWLS} model and uses the same \code{WLSCOMPONENT}, \code{WLSABSLENGTH}, and
\code{WLSTIMECONSTANT} material property interface. 

Recombination, quenching, or absolute scintillation yield models upstream of the optical transport, such as NEST, Birks quenching, or Modified Box  recombination~\cite{nest2011,nest2021} are not included in the current optical model. Our goal is to test optical transport, WLS absorption and re-emission, boundary handling, and detected hit production for a fixed source description.

\subsection{Photon source and GPU offloading workflow}
\label{sec:source}

The interface between the host \Geant{} simulation and the GPU optical transport is the \Opticks{} generation step (genstep). A genstep is a compact record for each \Geant{} step that encodes optical photon generation. It stores the
source position, source time, step direction, parent \Geant{} track identifier, and the number of optical photons assigned to that step by the native \Geant{} scintillation and Cerenkov processes. A single genstep typically expands to tens or occasionally thousands of optical photons in the GPU
photon generation kernel.

For the validation studies, the CPU and GPU transports are compared using the same recorded genstep input from a given primary event.  This isolates the optical transport stage from the upstream particle cascade and photon source generation. The two simulations therefore start from identical source records, while the photon generation and transport are performed by
the corresponding CPU or GPU optical stack.

The primary benchmark sample is a \SI{2.5}{GeV} electron generated at the centre of the liquid argon volume. This sample provides a compact,
high statistics electromagnetic shower topology with a photon load representative of a GeV-scale LArTPC event. In the validation and timing
runs, the \Geant{} electromagnetic cascade produces $61.0 \pm 0.2$ million optical photons from scintillation and Cerenkov processes. The Cerenkov contribution is less than \SI{0.05}{\percent} of the total generated photon count.

Two additional primary particles are used to test different photon source topologies. A contained muon sample provides a track like source distinct from the electromagnetic shower, while a \SI{400}{MeV} proton sample showcases a scintillation dominated, highly ionising topology. At this kinetic energy the
proton is close to the Cerenkov threshold in liquid argon, any Cerenkov component is negligible compared to the scintillation yield.

The same genstep based interface can also be used to attach event level and hit level truth information to the returned GPU hits. The present paper reports transported photon and hit validation, the MC truth association method and its overhead are described separately in
section~\ref{sec:mctruth}.

\subsection{Software and hardware configuration}
\label{sec:setup}

The CPU reference simulation uses \Geant{} 11.3.2. The GPU simulation uses
\Simphony on top of \Opticks, CUDA, and NVIDIA OptiX. \Simphony descends from EIC-Opticks~\cite{galgoczi2025eicopticks}, which introduced event aggregation and deployment guardrails for \Opticks-based optical transport. The benchmark GPU is an NVIDIA RTX 4090 with \SI{24}{GiB} of device memory, and the CPU reference is one Intel Xeon w7-3445 thread.

All headline timing comparisons in this paper use one GPU and one CPU thread. This choice provides a useful per thread reference for the algorithmic optical transport speedup. The software and hardware configuration used for the benchmarks is summarized in table~\ref{tab:software}.

\begin{table}[t]
  \centering
  \caption{Benchmark software and hardware configuration.}
  \label{tab:software}
  \begin{tabular}{ll}
    \toprule
    Item & Configuration \\
    \midrule
    Reference transport & \Geant{} 11.3.2, single CPU thread \\
    GPU transport & \Simphony built on \Opticks \\
    GPU & NVIDIA RTX 4090, \SI{24}{GiB} \\
    CPU reference & Intel Xeon w7-3445, one thread, 3.4 GHz \\
    CUDA & 13.0 \\
    OptiX & 9.0 \\
    NVIDIA driver & 580.105.08 \\
    \bottomrule
  \end{tabular}
\end{table}

\section{Wavelength shifting implementation}
\label{sec:wls}

Wavelength shifting is treated in \Simphony as a bulk optical transport process. This is important for liquid argon detector studies, where the primary scintillation light is produced in vacuum ultraviolet
near \SI{128}{nm} and must be converted to longer wavelengths before it can
be efficiently detected by most photon detectors. In the benchmark studied here, the process is exercised by a two stage wavelength-shifting chain: VUV photons are first shifted by the pTP layer and are then shifted
again by the TPB-doped acrylic layer before reaching the idealized photon detector.

A material is treated as wavelength shifting if its \Geant{} material property table provides a \code{WLSCOMPONENT} emission spectrum. The optional \code{WLSABSLENGTH} and \code{WLSTIMECONSTANT} entries set the absorption length and re-emission time constant if present, if \code{WLSTIMECONSTANT} is absent the re-emission is instantaneous. These are the same material property keys used by the \Geant{} \code{G4OpWLS} process. The GPU implementation therefore uses the same material property interface as the CPU reference simulation. The intent is not to require bitwise identical CPU and GPU execution, but to ensure that both simulations are driven by the same optical input data and by the same physical model.

For each WLS material, the tabulated \code{WLSCOMPONENT} emission spectrum is converted on the host into a normalized cumulative distribution function (CDF). The inverse cumulative distribution is then tabulated and copied to the GPU. The central inverse CDF table uses 4096 uniformly spaced cumulative probability nodes. Additional higher density tabulations are used in the lower and upper 5\% of the cumulative distribution to reduce interpolation residuals in the spectral tails.  The host side construction follows the same \Geant{} material property interpolation convention used for the reference simulation, so that the CPU and GPU emission samplers are based on a common energy grid. A per material integer index selects the appropriate WLS
emission table on the GPU, non-WLS materials are assigned an invalid value and cannot enter the WLS branch.

During transport through a WLS material, wavelength shifting competes with the other bulk processes. At each step, \Simphony samples candidate distances for ordinary absorption, Rayleigh scattering, and WLS absorption. For a photon of energy $E$, the WLS absorption distance is sampled as
\begin{equation}
  d_{\rm WLS} = -\lambda_{\rm WLS}(E)\ln \xi ,
\end{equation}
where $\lambda_{\rm WLS}(E)$ is obtained from the tabulated \code{WLSABSLENGTH} and $\xi$ is a uniform random number. The shortest candidate distance, together with any geometry boundary limitation, determines the next process. Ordinary absorption terminates the photon, Rayleigh scattering changes its direction according to the relevant optical model, WLS absorption moves the photon to the absorption point and re-emits it with a new wavelength, direction, polarization, and time.

At a WLS absorption point, the re-emitted photon energy is sampled from the material specific inverse CDF. The Stokes condition is enforced by requiring the re-emitted photon energy not to exceed the absorbed photon energy. If a sample violates this condition, the emission energy is resampled. The retry budget is capped at 100 iterations. For the spectra used in this benchmark, the loop terminates within a few iterations for all photons. The re-emission
direction is sampled isotropically. The new polarization is generated in the plane perpendicular to the new momentum using a trigonometric construction, which avoids the near parallel singularity that can occur in random cross product methods.

The WLS time delay is sampled from an exponential distribution with mean \code{WLSTIMECONSTANT},
\begin{equation}
  \Delta t_{\rm WLS} = -\tau_{\rm WLS}\ln \xi ,
\end{equation}
where $\tau_{\rm WLS}$ is the material specific WLS decay time. The same WLS time constants are used in the CPU and GPU configurations for the comparisons reported in this paper. The WLS photon multiplicity is one throughout this study, matching the single photon WLS configuration used by the reference material tables.

The GPU implementation updates the photon state in place after WLS absorption and records the WLS interaction in the photon process mask. It does not store a full per step optical history for every transported photon, because doing so would increase memory traffic and host--device transfer costs substantially. Instead, compact process flags and the detected hit record are retained for validation and analysis. This design is adequate for the benchmark comparisons in section~\ref{sec:validation}, where the relevant observables are hit counts, spectra, arrival times, positions,
and genstep level truth associations.

\section{Monte Carlo truth propagation}
\label{sec:mctruth}

For detector design studies and for machine learning training sample generation, it is often not sufficient to know only that an optical photon was detected. The detected hit should also be associated with the energy deposition step, event, or parent \Geant{} track that produced the photon. At the same time, carrying a full \Geant{} optical photon track object, or storing a complete per step history, would be impractical for events containing tens of millions of optical photons. \Simphony therefore retains compact genstep level truth information and reconstructs the source association only for photons that are actually detected.

If a genstep $g$ produces $p_g$ photons, its
photons are assigned a contiguous range of global photon indices beginning at the cumulative offset
\begin{equation}
  o_g = \sum_{h<g} p_h .
\end{equation}
A photon with global index $i$ is therefore associated with the unique genstep
$g$ satisfying
\begin{equation}
  o_g \leq i < o_g + p_g .
  \label{eq:truth-offset}
\end{equation}

Each detected hit returned from the GPU contains the photon position, time, direction, polarization, wavelength, packed boundary and material flags, an accumulated process mask, and the global photon index.  It does not contain a full \Geant{} track object. The source genstep is recovered on the host by a predecessor query on the sorted cumulative offset array. Once the genstep is identified, the hit can be labeled with the corresponding event, source position and time, parent \Geant{} track identifier, and any additional
genstep-level metadata recorded by the upstream simulation.

This association is exact at the genstep level. It identifies which \Geant{} source step generated the detected photon, but it does not reconstruct the full optical path taken by that photon. Information about the transport history is limited to the compact process and boundary flags stored in the hit record. More detailed optical histories could be recorded for diagnostic runs, but they are not used in the  workflow studied here.

For the \SI{2.5}{GeV} electron benchmark, a typical event launches approximately \num{6.08e7} optical photons from about \num{6.7e4} gensteps and produces about \num{2.4e6} detected hits at the outer photon counting boundary. The detected fraction is therefore approximately \SI{3.9}{\percent}.  The host side source lookup requires
$\log_2(\num{6.7e4}) \simeq 16$ comparisons per detected hit. In the current implementation this corresponds to about \SI{80}{ns} per detected hit, including the offset-array access. The total truth association cost is then
approximately
\begin{equation}
  T_{\rm truth}
  \simeq
  N_{\rm hit}\,t_{\rm lookup}
  \simeq
  \num{2.4e6}\times\SI{80}{ns}
  \simeq
  \SI{0.2}{s}
\end{equation}
per-event.  Expressed per launched photon, the cost is
\begin{equation}
  \frac{T_{\rm truth}}{N_{\rm phot}}
  \simeq
  f_{\rm det}\,t_{\rm lookup}
  \simeq
  0.039\times\SI{80}{ns}
  \simeq
  \SI{3.1}{ns}.
\end{equation}

The truth association overhead is therefore amortized over the much larger population of transported photons that are not detected. For the \SI{14.7}{kt} benchmark operating point, it is roughly \SI{25}{\percent} of the per-photon GPU optical transport cost and about \SI{2}{\percent} of the full end-to-end event wall time. This scaling is favorable for large
liquid noble detectors, where the launched photon multiplicity is high but the detected fraction is typically small.

\section{Validation against \Geant}
\label{sec:validation}

\subsection{Validation strategy and statistical comparison}
\label{sec:validation-method}

The validation compares \Simphony with the \Geant{} 11.3.2 optical photon reference simulation in the benchmark geometry described in section~\ref{sec:simulation-model}. For each primary particle configuration, the CPU and GPU transports are run from the same optical photon source distributions. This use of identical genstep input isolates the optical transport, wavelength shifting, optical boundary handling, and detected hit formation from the EM shower of the creating particles.

The first comparison metric is the integrated detected photon ratio,
\begin{equation}
  R_N = \frac{N_{\rm Simphony}}{N_{\rm Geant}},
  \label{eq:rn}
\end{equation}
where $N_{\rm Simphony}$ and $N_{\rm Geant}$ are the numbers of photons
recorded at the idealized photon detector. Unless otherwise stated, the quoted uncertainties on integrated hit counts and on $R_N$ are the sample standard deviations over three paired random seed runs.

For one dimensional spectra, the comparison uses the binned statistic
\begin{equation}
  \frac{\chi^2}{\mathrm{ndf}}
  =
  \frac{1}{N_{\rm occ}}
  \sum_{i\in{\rm occ}}
  \frac{\left(S_i-G_i\right)^2}{S_i+G_i},
  \label{eq:chi2}
\end{equation}
where $S_i$ and $G_i$ are the \Simphony and \Geant{} bin counts, and the sum
runs over the $N_{\rm occ}$ bins for which $S_i+G_i>0$. The histograms are
not normalized before forming equation~\ref{eq:chi2}. The arrival time sum runs over $N_{\rm occ}\approx995$ non zero bins
  (0--\SI{5000}{ns} in \SI{5}{ns} steps) and the wavelength sum over
  $N_{\rm occ}\approx166$ (\SI{60}{nm}--\SI{820}{nm} in \SI{2}{nm} steps).

No single metric is standard across published GPU optical photon validation studies. The original \Opticks{} validation for JUNO emphasized random number aligned photon history comparisons~\cite{blyth2019opticks}, the EIC pfRICH validation reported integrated hit count agreement at the $4\times10^{-5}$ level~\cite{galgoczi2025eicopticks}, and the NEXT-CRAB-0 GPU study compared channel level Gaussian fit means at the percent level~\cite{nextcrab2025}. The present validation therefore reports both integrated hit ratios and binned spectral or spatial comparisons.

\subsection{Electron shower validation}
\label{sec:electron-validation}

The primary validation sample is a \SI{2.5}{GeV} electron generated at the centre of the liquid argon volume. This sample produces a compact electromagnetic shower and approximately \SI{61}{M} optical photons per
event. It is therefore a high statistics test of bulk propagation, two stage wavelength-shifting, boundary handling, and hit formation.

Table~\ref{tab:val-electron} summarizes the comparison. Averaged over three
paired random seed runs, the integrated hit ratio is
\[
  R_N = 1.0022 \pm 0.0002 .
\]
The central value differs from unity by about \SI{0.22}{\percent}, well below
the percent level. The detected arrival time and arrival wavelength spectra give
$\chisqndf=0.98$ and $\chisqndf=1.08$, respectively, on the seed-42
single event histograms.

\begin{table}[t]
  \centering
  \footnotesize
  \setlength{\tabcolsep}{3pt}
  \renewcommand{\arraystretch}{1.15}
  \caption{Validation for the \SI{2.5}{GeV} electron sample at the center of the benchmark geometry. Each event contains approximately \SI{61}{M} generated optical photons. Hit counts are reported in units of $10^6$ and are given as mean $\pm$ one standard deviation over three paired \Geant{}+\Simphony{} random seed simulations. Spectral $\chisqndf$ values are evaluated on the seed 42 single event histograms.}
  \label{tab:val-electron}
  \begin{tabularx}{\linewidth}{@{}
    >{\raggedright\arraybackslash}X
    >{\centering\arraybackslash}p{0.19\linewidth}
    >{\centering\arraybackslash}p{0.19\linewidth}
    >{\centering\arraybackslash}p{0.22\linewidth}
    @{}}
    \toprule
    Observable & \Geant{} & \Simphony{} & Result \\
    \midrule
    Detected photons ($10^6$)
      & \num{2.378(55)}
      & \num{2.383(55)}
      & $R_N = 1.0022 \pm 0.0002$ \\
    Arrival time spectrum,
    0--\SI{5000}{ns}, \SI{5}{ns} bins
      & --
      & --
      & $\chisqndf=0.98$ \\
    Arrival wavelength spectrum,
    \SI{60}{nm}--\SI{820}{nm}, \SI{2}{nm} bins
      & --
      & --
      & $\chisqndf=1.08$ \\
    \bottomrule
  \end{tabularx}
\end{table}

Across 20 independent \Simphony{} launches with the same nominal primary configuration, the hit count is $(2.37\pm0.08)\times10^{6}$, corresponding to a fractional spread of
\SI{3.5}{\percent}. This seed-to-seed variation reflects the combined fluctuations of the source generation and optical transport for the primary configuration. The paired \Simphony/\Geant{} ratios remain stable within the quoted statistical uncertainties.

\begin{figure}[t]
  \centering
  \includegraphics[width=0.78\linewidth]{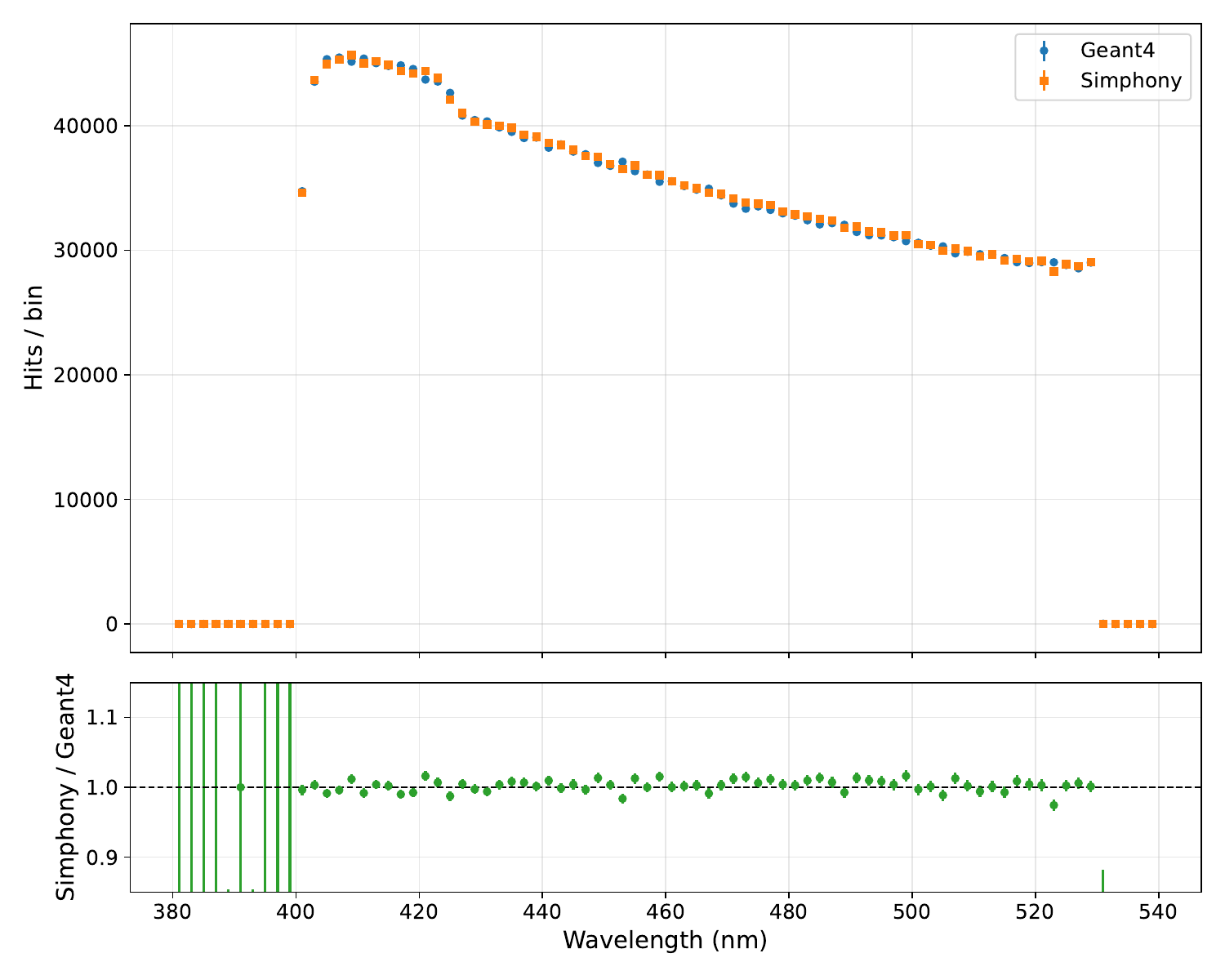}
  \caption{Detected wavelength spectrum for one \SI{2.5}{GeV} electron event. The comparison uses the same seed 42 source gensteps in the \Geant{} and
  \Simphony{} optical transports.}
  \label{fig:2p5-wavelength}
\end{figure}

The detected wavelength spectrum is shown in figure~\ref{fig:2p5-wavelength}. It is dominated by the second WLS emission band near \SI{425}{nm}. The intermediate pTP emission near \SI{340}{nm} is absorbed and re-emitted in the wavelength-shifting acrylic in both
simulations. 

\begin{figure}[t]
  \centering
  \includegraphics[width=0.78\linewidth]{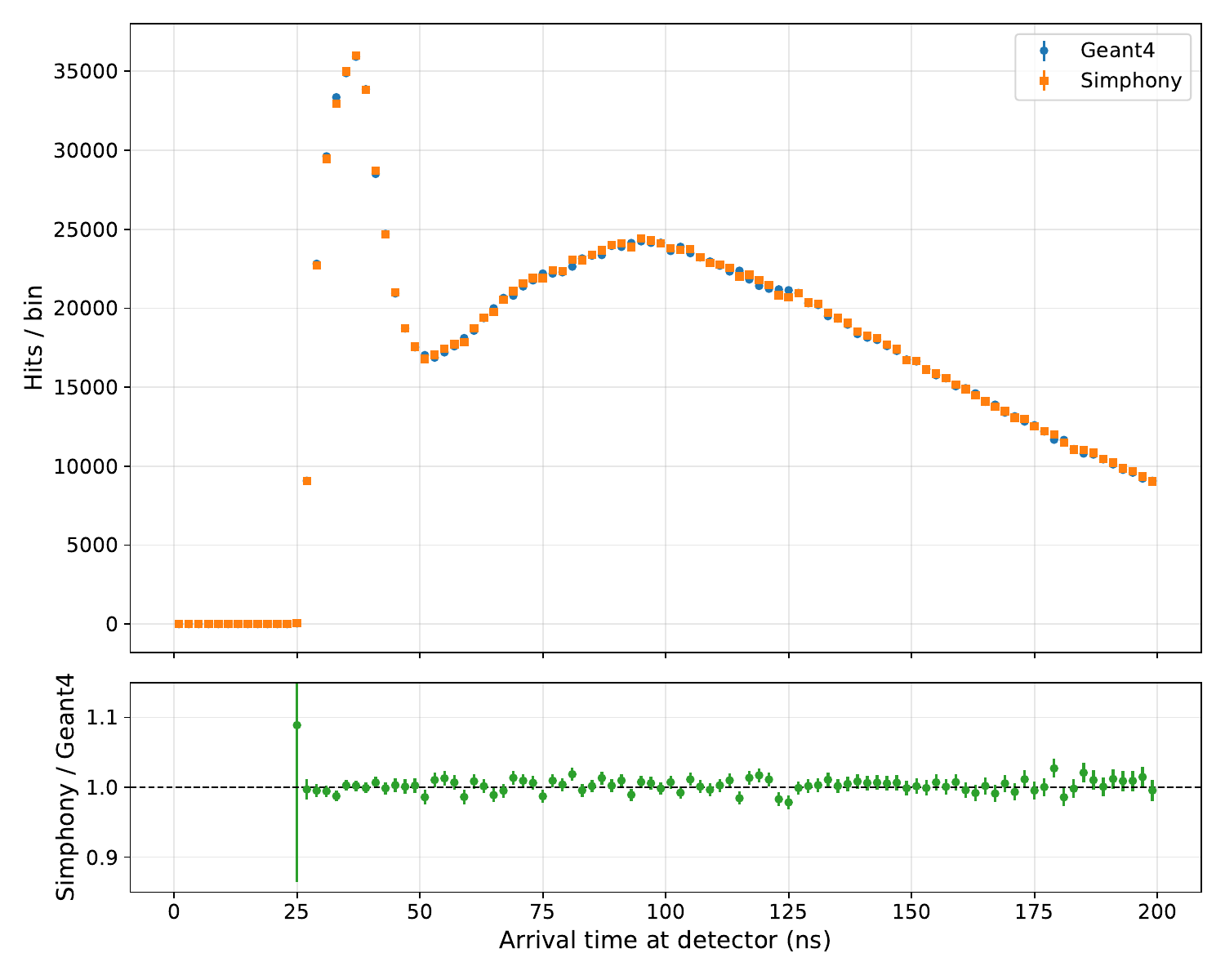}
  \caption{Detected arrival time spectrum for one \SI{2.5}{GeV} electron
  event. The histogram includes both prompt photons and the slow
  scintillation component.}
  \label{fig:2p5-arrival}
\end{figure}

The detected arrival time distribution is shown in figure~\ref{fig:2p5-arrival}. The spectrum contains both the prompt light and the long slow scintillation tail. Agreement in this distribution is a sensitive test of the WLS time sampling, group velocity, Rayleigh scattering, and optical boundary treatment.  Because the source gensteps are identical in the two transports, the comparison tests the optical transport implementation rather than the upstream absolute scintillation yield model.

\begin{figure}[t]
  \centering
  \includegraphics[width=0.95\linewidth]{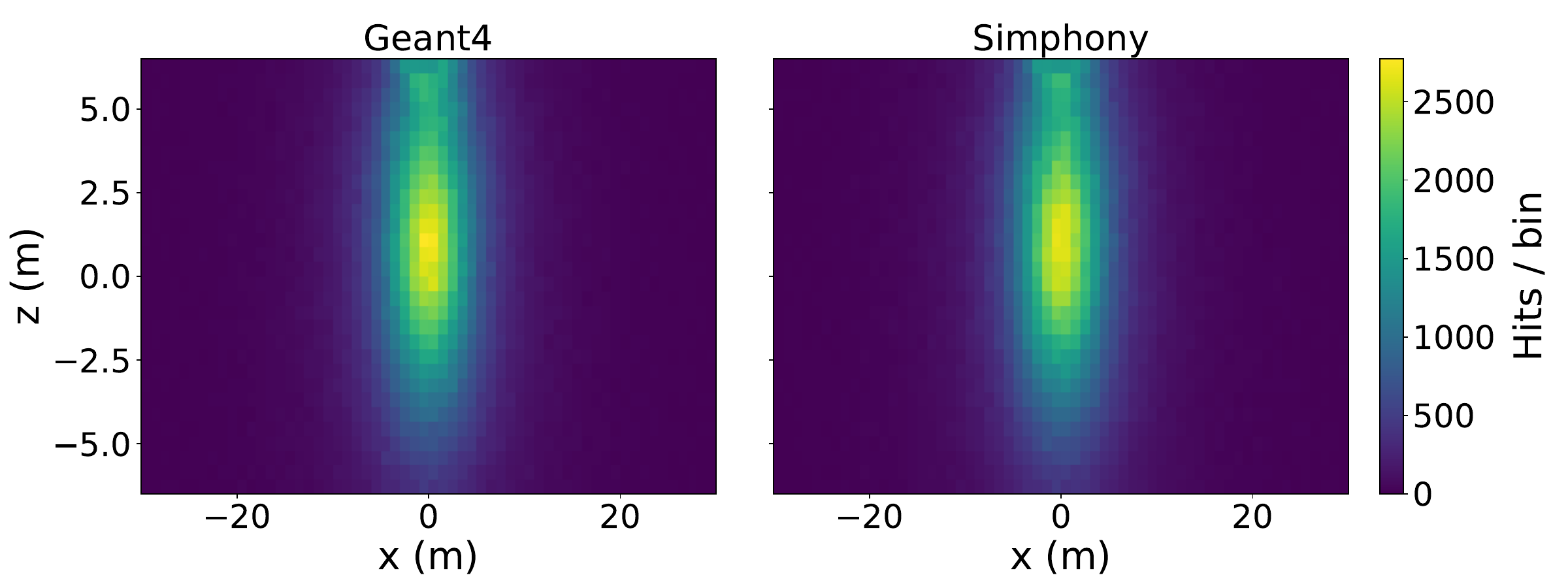}
  \caption{Two dimensional hit maps on the $+Y$ detector face for one \SI{2.5}{GeV} electron event. The left panel shows the \Geant{} reference result and the right panel shows the \Simphony{} result.}
  \label{fig:2p5-hitmap}
\end{figure}

Figure~\ref{fig:2p5-hitmap} compares the two dimensional hit distributions on the $+Y$ face of the photon detector. The forward asymmetric distribution associated with the primary particle direction is reproduced by the GPU transport.

\begin{figure}[t]
  \centering
  \includegraphics[width=0.85\linewidth]{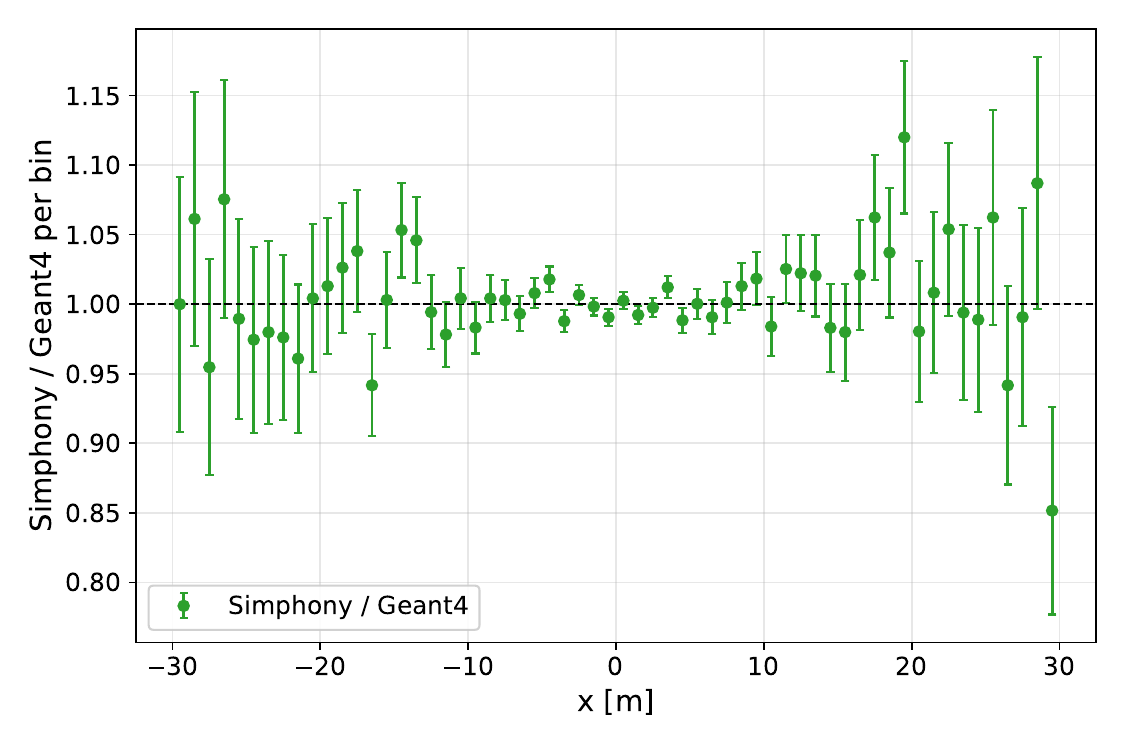}
  \caption{Per-bin ratio $N_\mathrm{Simphony}/N_\mathrm{Geant}$ on the  $+Y$ face for the seed 42 \SI{2.5}{GeV} electron event, projected onto the long $x$ axis. Errorbars are the Poisson propagated 1$\sigma$
  uncertainties from equation~\ref{eq:ratio-error}. The dashed line marks
  unit ratio. The unit ratio test yields $\chisqndf=0.82$.}
  \label{fig:2p5-ratio-x}
\end{figure}

The spatial agreement is quantified by projecting the $+Y$-face hit map onto the two in plane coordinates and forming the per-bin ratio $N_{\rm Simphony}/N_{\rm Geant}$.  The uncertainty on each ratio is computed from independent Poisson statistics in the two samples,
\begin{equation}
  \frac{\sigma_R}{R}
  =
  \sqrt{\frac{1}{N_{\rm Simphony}}+\frac{1}{N_{\rm Geant}}}.
  \label{eq:ratio-error}
\end{equation}
Figure~\ref{fig:2p5-ratio-x} shows the projection along the \SI{60}{m} $x$ axis using 60 bins of \SI{1}{m}. Testing the ratios against unity yields $\chisqndf=0.82$ for the $x$ projection and $\chisqndf=1.31$ for the $z$ projection, consistent with bin-by-bin
statistical agreement.

\subsection{Muon and proton validation samples}
\label{sec:multipart-validation}

The electron shower test is supplemented by two additional source topologies. A contained \SI{1}{GeV} muon provides a track like optical source, while a \SI{400}{MeV} proton provides a lower yield, highly ionizing topology. For the refractive index used in this benchmark, the proton is close to the Cerenkov threshold in liquid argon, so its optical signal is dominated by scintillation.

The results are summarized in table~\ref{tab:multipart}. The electron row is repeated for comparison. The three integrated ratios lie between 1.0005 and 1.0022, corresponding to agreement better than \SI{0.25}{\percent} in the central values for all three primary samples. The time and wavelength
$\chisqndf$ values are close to unity for the stated histogram binning.

\begin{table}[t]
  \centering
  \footnotesize
  \setlength{\tabcolsep}{3pt}
  \renewcommand{\arraystretch}{1.15}
  \caption{Validation results in the benchmark geometry. Hit counts
  are reported in units of $10^6$ and are given as mean $\pm$ one standard deviation over three paired \Geant{}+\Simphony{} random-seed runs. Spectral $\chisqndf$ values are evaluated on the seed 42 single event
  histograms. The electron row is repeated from table~\ref{tab:val-electron}
  for comparison.}
  \label{tab:multipart}
  \begin{tabularx}{\linewidth}{@{}
    >{\raggedright\arraybackslash}p{0.08\linewidth}
    >{\centering\arraybackslash}p{0.12\linewidth}
    >{\centering\arraybackslash}p{0.16\linewidth}
    >{\centering\arraybackslash}p{0.16\linewidth}
    >{\centering\arraybackslash}p{0.22\linewidth}
    >{\centering\arraybackslash}X
    @{}}
    \toprule
    Primary
      & \shortstack{Kinetic\\energy}
      & \shortstack{\Geant{} hits\\($10^6$)}
      & \shortstack{\Simphony{} hits\\($10^6$)}
      & $R_N$
      & \shortstack{Time / wavelength\\$\chisqndf$} \\
    \midrule
    $e^-$
      & \SI{2.5}{GeV}
      & \num{2.378(55)}
      & \num{2.383(55)}
      & $1.0022 \pm 0.0002$
      & 0.98 / 1.08 \\
    $\mu^-$
      & \SI{1.0}{GeV}
      & \num{1.421(34)}
      & \num{1.423(34)}
      & $1.0017 \pm 0.0008$
      & 1.01 / 0.90 \\
    $p$
      & \SI{400}{MeV}
      & \num{0.259(25)}
      & \num{0.259(24)}
      & $1.0005 \pm 0.0014$
      & 1.07 / 1.03 \\
    \bottomrule
  \end{tabularx}
\end{table}

Across the 20 launch \Simphony{} performance samples, the hit counts are $(2.37\pm0.08)\times10^{6}$ for the electron,
$(1.44\pm0.05)\times10^{6}$ for the muon, and
$(2.64\pm0.24)\times10^{5}$ for the proton.  The corresponding fractional spreads are \SI{3.5}{\percent}, \SI{3.3}{\percent}, and
\SI{9.2}{\percent}.

\begin{figure}[t]
  \centering
  \begin{subfigure}[b]{0.48\linewidth}
    \includegraphics[width=\linewidth]{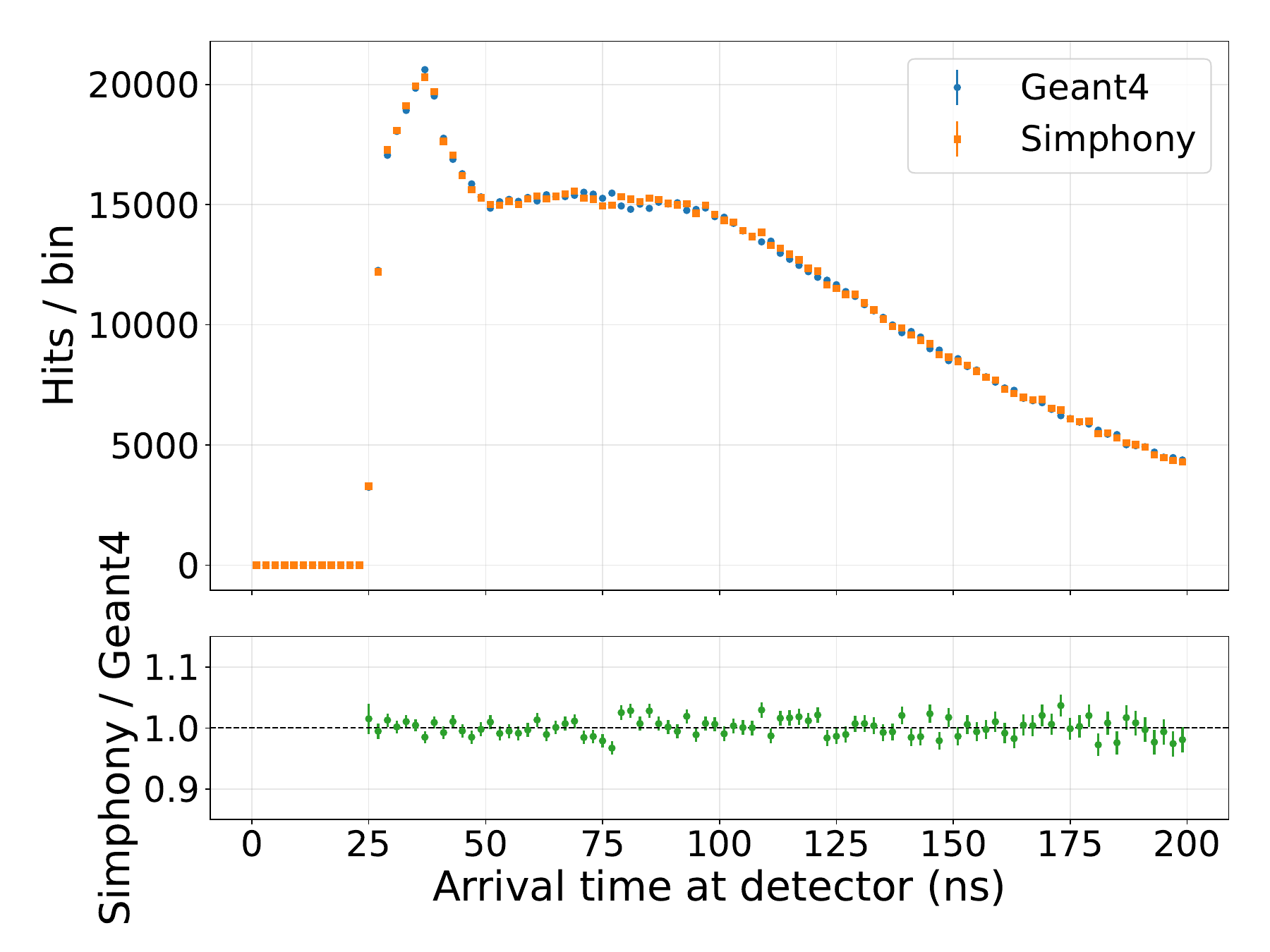}
    \caption{\SI{1}{GeV} muon.}
    \label{fig:muon-time}
  \end{subfigure}
  \hfill
  \begin{subfigure}[b]{0.48\linewidth}
    \includegraphics[width=\linewidth]{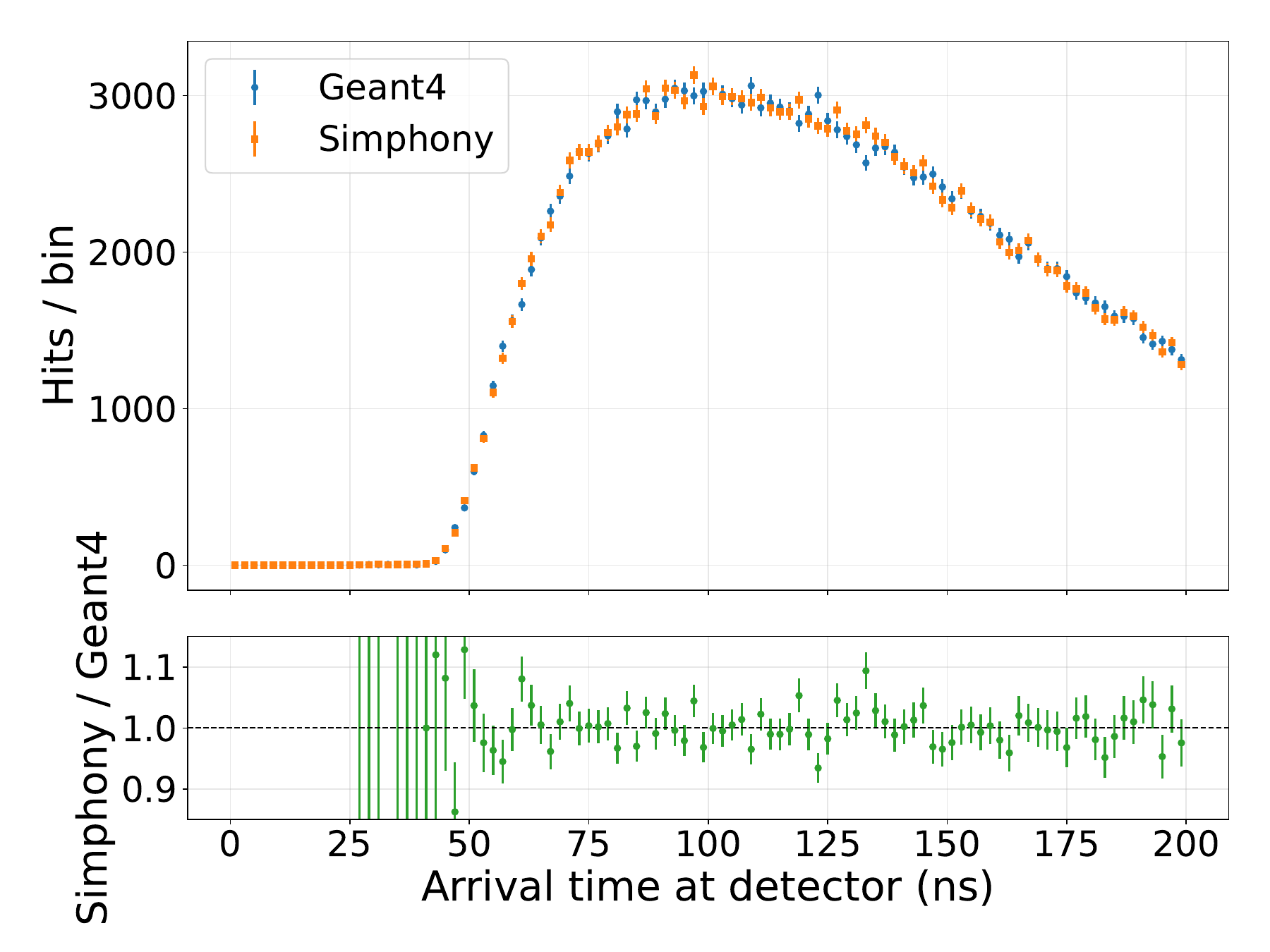}
    \caption{\SI{400}{MeV} proton.}
    \label{fig:proton-time}
  \end{subfigure}
  \caption{Detected arrival time spectra for the muon and proton validation
  samples.}
  \label{fig:multipart-time}
\end{figure}

Figure~\ref{fig:multipart-time} shows the detected arrival time spectra for the muon and proton samples. The agreement is comparable to that observed in the electron shower benchmark, despite the different source geometries and
photon yields.

\begin{figure}[t]
  \centering
  \begin{subfigure}[b]{0.95\linewidth}
    \includegraphics[width=\linewidth]{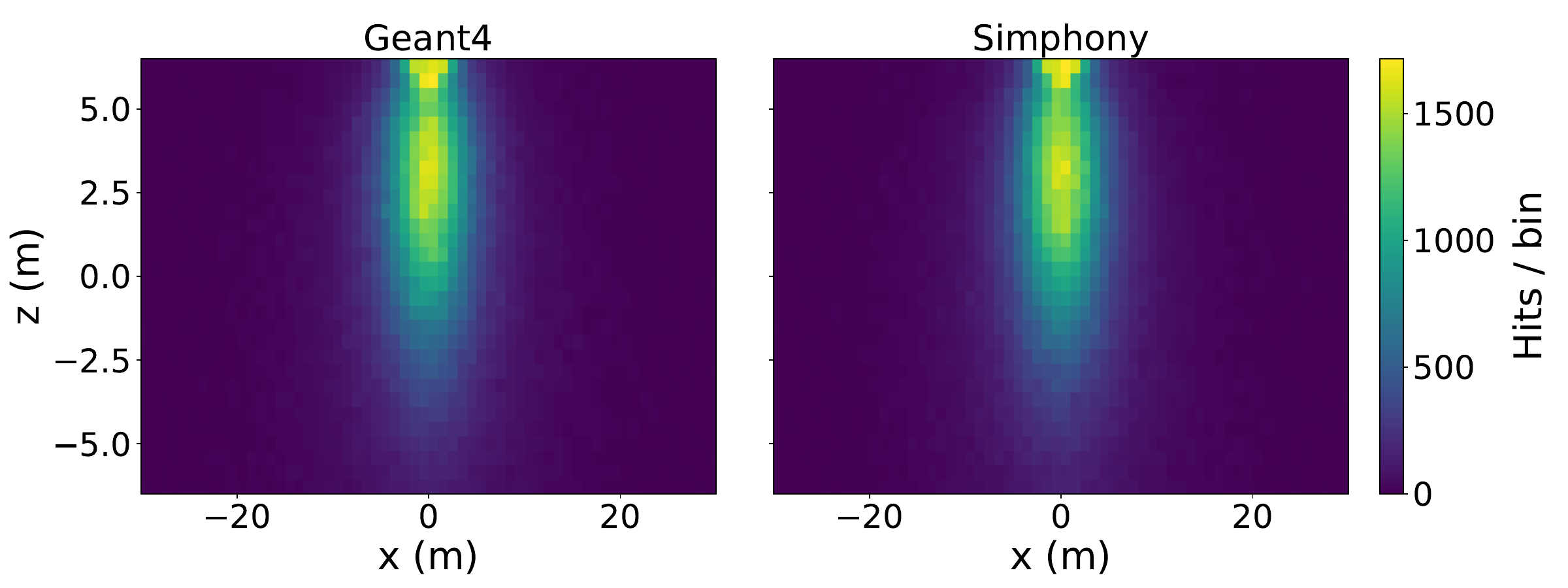}
    \caption{\SI{1}{GeV} muon sample.}
    \label{fig:muon-hitmap}
  \end{subfigure}\\[1ex]
  \begin{subfigure}[b]{0.95\linewidth}
    \includegraphics[width=\linewidth]{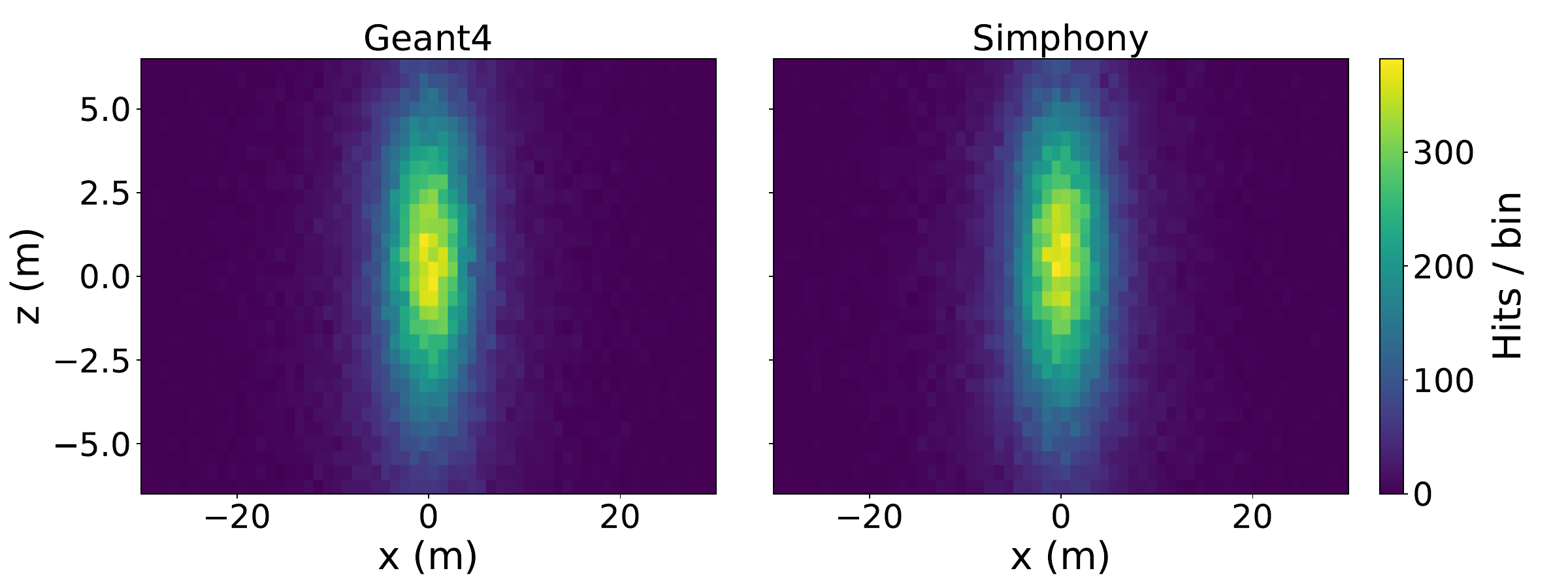}
    \caption{\SI{400}{MeV} proton sample.}
    \label{fig:proton-hitmap}
  \end{subfigure}
  \caption{Two dimensional hit maps on the $+Y$ detector face for the muon and proton validation samples.}
  \label{fig:multipart-hitmaps}
\end{figure}

The corresponding $+Y$-face hit maps are shown in figure~\ref{fig:multipart-hitmaps}. The muon sample gives a track like spatial response, while the proton sample gives a lower yield and more compact response. In both cases, the GPU transport reproduces the \Geant{} spatial pattern. The one-dimensional projection ratio checks for these samples are given in appendix~\ref{sec:appendix-ratio}. The corresponding $\chisqndf$ values are 0.92 and 1.32 for the muon $x$ and $z$ projections, and 0.80 and 0.75 for the proton $x$ and $z$ projections.

Under the shared genstep comparison, the electron, muon, and proton samples show subpercent integrated agreement and statistically consistent timing, wavelength, and spatial distributions.

\section{Performance}
\label{sec:performance}

\subsection{Timing definitions and throughput}
\label{sec:timing-definitions}

We define three timing quantities. The first is the single-thread \Geant{} optical reference time, $T_{\mathrm{G4,opt}}$. This is obtained from paired \Geant{} runs with optical photon transport enabled and disabled, the difference of the two wall times isolates the CPU optical transport stage from the upstream particle
cascade and energy deposition calculation.  This reference is useful for measuring the algorithmic acceleration of optical transport.

The second quantity is the GPU optical interval, $T_{\mathrm{GPU,opt}}$. It includes the host-to-device transfer of genstep records, the GPU optical propagation and hit compaction, and the device-to-host return of compacted detected hits. It does not include the upstream \Geant{} particle simulation, geometry setup outside the optical launch or the output writing.

The third quantity is the end-to-end wall time,
$T_{\mathrm{e2e}}$. This includes the host \Geant{} particle simulation, geometry and acceleration structure setup, GPU optical transport, hit return, output writing, and process teardown. The corresponding end-to-end speedup reported below is defined with respect to the same single thread \Geant{} optical reference, $T_{\mathrm{G4,opt}}$. 

For the \SI{2.5}{GeV} electron benchmark, the single-thread \Geant{} optical
reference is
\[
  T_{\mathrm{G4,opt}} = 798.6 \pm 39.1~\mathrm{s}
\]
per-event.  A single GPU launch transporting one such event takes $0.97 \pm 0.02$ s.  Stacking four electron events into one GPU launch transports $243 \pm 3$ million photons in $3.03 \pm 0.06$ s, corresponding to a sustained optical transport throughput of $80.2 \pm 1.6$ million photons s$^{-1}$. The resulting optical transport speedup is $1053 \pm 55$ relative to the single-thread \Geant{} reference, including genstep upload and hit return.

The same benchmark was repeated for a contained \SI{1}{GeV} muon and for a \SI{400}{MeV} proton. The corresponding single thread \Geant{} optical reference times are $432.1 \pm 19.3$ s and $89.6 \pm 9.9$ s per-event, respectively.  Table~\ref{tab:speedup-points} summarizes the GPU optical timing for the three primary samples. The stacked launch multiplicities were chosen to use a large fraction of the RTX 4090 memory while remaining below the \SI{24}{GiB} device memory limit.

\begin{table}[t]
  \centering
  \footnotesize
  \setlength{\tabcolsep}{4pt}
  \renewcommand{\arraystretch}{1.12}
  \caption{GPU optical transport timing for the benchmark samples. The GPU optical time includes host-to-device genstep upload, GPU optical propagation and hit compaction, and device-to-host hit return. It excludes the upstream \Geant{} particle simulation and output I/O. Values are mean $\pm$ one standard deviation over 20 random seed launches. The optical speedup is defined as
  $N T_{\mathrm{G4,opt}}/T_{\mathrm{GPU,opt}}$, where $N$ is the number of events in the stacked launch. The single thread \Geant{} optical references are $798.6 \pm 39.1$ s, $432.1 \pm 19.3$ s, and $89.6 \pm 9.9$ s per-event for the electron, muon, and proton samples, respectively.}
  \label{tab:speedup-points}
  \begin{tabular}{@{}
    >{\raggedright\arraybackslash}p{0.20\linewidth}
    c
    c
    c
    c
    c
    @{}}
    \toprule
    Sample
      & Events/launch
      & Photons/launch
      & Hits/launch
      & GPU optical time
      & Optical speedup \\
    \midrule
    $e^-$, \SI{2.5}{GeV}
      & 1
      & \num{6.10(2)e7}
      & \num{2.37(8)e6}
      & $0.97 \pm 0.02$ s
      & $822 \pm 43$ \\
    $e^-$, \SI{2.5}{GeV}
      & 4
      & \num{2.43(3)e8}
      & \num{9.47(14)e6}
      & $3.03 \pm 0.06$ s
      & $1053 \pm 55$ \\
    $\mu^-$, \SI{1}{GeV}
      & 1
      & \num{2.48(3)e7}
      & \num{1.44(5)e6}
      & $0.79 \pm 0.02$ s
      & $551 \pm 27$ \\
    $\mu^-$, \SI{1}{GeV}
      & 7
      & \num{1.749(9)e8}
      & \num{1.011(16)e7}
      & $2.54 \pm 0.05$ s
      & $1190 \pm 59$ \\
    $p$, \SI{400}{MeV}
      & 1
      & \num{8.0(16)e6}
      & \num{2.64(24)e5}
      & $0.39 \pm 0.05$ s
      & $231 \pm 40$ \\
    $p$, \SI{400}{MeV}
      & 23
      & \num{1.89(6)e8}
      & \num{5.89(15)e6}
      & $2.15 \pm 0.04$ s
      & $959 \pm 108$ \\
    \bottomrule
  \end{tabular}
\end{table}

The optical speedup increases when events are stacked because fixed launch costs and other per launch overheads are amortised over more photons. This effect is most visible for the lower yield proton sample: a single \SI{400}{MeV} proton event yields an optical speedup of $231 \pm 40$, while a 23-event stacked launch yields $959 \pm 108$. The muon sample shows the same trend. For the electron sample the single event photon load is already large, so the additional gain from four event stacking is smaller but still visible.

Table~\ref{tab:e2e-memory} gives the corresponding end-to-end wall time and peak GPU memory usage. For the four event \SI{2.5}{GeV} electron stack, the full benchmark process takes $36.0 \pm 0.8$ s, or $9.00 \pm 0.20$ s per-event.  This gives an end-to-end acceleration factor of $89 \pm 5$ relative to the same single thread \Geant{} optical reference. The peak GPU memory use is approximately \SI{19.7}{GiB}, or about \SI{82}{\percent} of the RTX 4090 device memory.  Larger electron batches must therefore be
split across multiple launches on this GPU.

\begin{table}[t]
  \centering
  \footnotesize
  \setlength{\tabcolsep}{4pt}
  \renewcommand{\arraystretch}{1.12}
  \caption{End-to-end wall time and peak GPU memory for the operating points   in table~\ref{tab:speedup-points}. The end-to-end time per-event is the total wall time divided by the stacked event multiplicity.  The end-to-end speedup is defined as $T_{\mathrm{G4,opt}}/T_{\mathrm{e2e}}$ per-event, using the same single thread \Geant{} optical reference as table~\ref{tab:speedup-points}.}
  \label{tab:e2e-memory}
  \begin{tabular}{@{}
    >{\raggedright\arraybackslash}p{0.28\linewidth}
    c
    c
    c
    c
    @{}}
    \toprule
    Sample
      & Events/launch
      & E2E time/event
      & E2E speedup
      & Peak GPU memory \\
    \midrule
    $e^-$, \SI{2.5}{GeV}
      & 1
      & $9.68 \pm 0.41$ s
      & $83 \pm 5$
      & \SI{5.9}{GiB} \\
    $e^-$, \SI{2.5}{GeV}
      & 4
      & $9.00 \pm 0.20$ s
      & $89 \pm 5$
      & \SI{19.7}{GiB} \\
    $\mu^-$, \SI{1}{GeV}
      & 1
      & $6.29 \pm 0.22$ s
      & $69 \pm 4$
      & \SI{4.6}{GiB} \\
    $\mu^-$, \SI{1}{GeV}
      & 7
      & $5.07 \pm 0.12$ s
      & $85 \pm 4$
      & \SI{13}{GiB} \\
    $p$, \SI{400}{MeV}
      & 1
      & $2.05 \pm 0.12$ s
      & $44 \pm 6$
      & \SI{2.7}{GiB} \\
    $p$, \SI{400}{MeV}
      & 23
      & $0.96 \pm 0.04$ s
      & $94 \pm 11$
      & \SI{16}{GiB} \\
    \bottomrule
  \end{tabular}
\end{table}

For scale, the four event electron throughput corresponds to approximately 13 minutes of GPU optical transport time for a 1000 event \SI{2.5}{GeV} electron sample on one RTX 4090.  Including the upstream host side simulation and benchmark output handling, the same sample requires approximately 2.5 hours end to end.  The single thread \Geant{} optical transport alone would require about 222 hours for the same number of events. At samples of $10^{5}$--$10^{6}$ events, the optical transport component therefore changes from years to decades of single thread CPU time to days to weeks of GPU optical transport time. However we note that a collaboration level production estimate must also include multi-threaded CPU baselines, production I/O, workflow scheduling, and the target experiment framework.

\subsection{Wall time decomposition and per-photon scaling}
\label{sec:timing-decomposition}

The benchmark wall time was further decomposed to identify the dominant workflow components and to provide a simple scaling model for other photon loads in the same geometry. Four contributions are separated:
\begin{itemize}
  \item the host side contribution, including the upstream \Geant{} particle
  simulation, geometry setup, and output writing
  \item the host-to-device transfer of genstep records
  \item the GPU propagation kernel, including on-device hit compaction
  \item the device-to-host transfer of compacted detected hits
\end{itemize}
With these definitions,
\begin{equation}
  T_{\mathrm{wall}}
  =
  T_{\mathrm{host}}
  +
  T_{\mathrm{HtoD}}
  +
  T_{\mathrm{kernel}}
  +
  T_{\mathrm{DtoH}} .
  \label{eq:wall-decomposition}
\end{equation}

This decomposition was measured in a separate controlled timing run for stacked launches of $N=1,2,3,4$ \SI{2.5}{GeV} electron events, with five random seeds at each stacking multiplicity. The same GPU RAM allocation was used for all measurements so that photon buffer pre-allocation did not change between operating points. Each primary produces approximately \num{61} million optical photons, so the timing scan covers approximately \num{61}--\num{243} million launched photons per GPU call. 

Table~\ref{tab:timing-decomposition} shows the measured component times. The host term is the largest contribution to the total wall time over the tested range. The genstep upload is negligible on this scale, while the returned hit data is small compared to both the host simulation and the GPU propagation kernel.

\begin{table}[t]
  \centering
  \footnotesize
  \setlength{\tabcolsep}{4pt}
  \renewcommand{\arraystretch}{1.12}
  \caption{Per launch wall time decomposition for stacked \SI{2.5}{GeV} electron events.  Values are mean $\pm$ one standard deviation over five random seed launches at each stacking multiplicity. The sum of the four components equals the wall time to the quoted
  precision.}
  \label{tab:timing-decomposition}
  \begin{tabular}{@{} c c c c c c c @{}}
    \toprule
    $N$
      & Photons
      & Host (s)
      & HtoD (ms)
      & Kernel (s)
      & DtoH (ms)
      & Wall (s) \\
    \midrule
    1
      & \num{60.8}\,M
      & $8.15 \pm 0.27$
      & $0.45$
      & $0.893 \pm 0.009$
      & $70 \pm 2$
      & $9.11 \pm 0.28$ \\
    2
      & \num{121.7}\,M
      & $16.56 \pm 0.22$
      & $0.85$
      & $1.464 \pm 0.027$
      & $137 \pm 4$
      & $18.17 \pm 0.24$ \\
    3
      & \num{182.6}\,M
      & $24.34 \pm 0.42$
      & $1.20$
      & $2.109 \pm 0.021$
      & $215 \pm 2$
      & $26.67 \pm 0.40$ \\
    4
      & \num{243.5}\,M
      & $33.35 \pm 0.64$
      & $1.70$
      & $2.754 \pm 0.030$
      & $280 \pm 6$
      & $36.39 \pm 0.67$ \\
    \bottomrule
  \end{tabular}
\end{table}

The component times are well described by linear functions of the launched photon count,
\begin{equation}
  T(N_{\mathrm{phot}})
  =
  \alpha + \gamma N_{\mathrm{phot}},
  \label{eq:component-linear-model}
\end{equation}
where $\alpha$ is a fixed per-launch contribution and $\gamma$ is the
per-photon slope. A weighted least squares fit was performed using the uncertainty on the mean at each stacking multiplicity, $\sigma_{\mathrm{mean}}=\sigma_{\mathrm{seeds}}/\sqrt{5}$.  The fitted parameters are shown in table~\ref{tab:timing-fit}.

\begin{table}[t]
  \centering
  \footnotesize
  \setlength{\tabcolsep}{6pt}
  \renewcommand{\arraystretch}{1.15}
  \caption{Linear fit parameters for
  $T(N_{\mathrm{phot}})=\alpha+\gamma N_{\mathrm{phot}}$ applied to the
  decomposition in table~\ref{tab:timing-decomposition}. The intercept
  $\alpha$ is the fixed per launch contribution and $\gamma$ is the per launched photon slope.}
  \label{tab:timing-fit}
  \begin{tabular}{@{} l c c @{}}
    \toprule
    Component
      & $\alpha$ (s)
      & $\gamma$ (ns / launched photon) \\
    \midrule
    Host: \Geant{} particle simulation, setup, I/O
      & $0.03 \pm 0.17$
      & $136.3 \pm 1.4$ \\
    HtoD: genstep upload
      & $\sim 0$
      & $0.01$ \\
    Kernel: OptiX propagation and compaction
      & $0.285 \pm 0.007$
      & $10.11 \pm 0.06$ \\
    DtoH: detected-hit download
      & $\sim 0$
      & $1.18 \pm 0.01$ \\
    \midrule
    \textbf{Wall total}
      & $\boldsymbol{0.31 \pm 0.18}$
      & $\boldsymbol{147.3 \pm 1.4}$ \\
    \bottomrule
  \end{tabular}
\end{table}

Several features of this decomposition are important for large noble liquid detector studies. First, after GPU acceleration of the optical transport, the dominant remaining cost in this benchmark is host-side work.  At the four event electron operating point, the host contribution is about \SI{92}{\percent} of the total wall time. The fitted host slope,
$136.3 \pm 1.4$ ns per launched photon, is approximately thirteen times the fitted GPU kernel slope. Thus, for this benchmark, further reductions in end-to-end wall time would require faster upstream event generation, improved host side workflow integration, or more aggressive event-level parallelism.

Second, the GPU kernel contains a fixed contribution of $0.285 \pm 0.007$ s. This term is about \SI{32}{\percent} of the kernel time for a single \SI{2.5}{GeV} electron event, but only about \SI{10}{\percent} for the four event stack. Event stacking is therefore an effective way to increase optical transport throughput, provided that the stacked photon and hit buffers remain within device memory.

Third, the transfer costs are small for this benchmark. The genstep upload stays below \SI{2}{ms} even for the four event electron stack.  The hit download is larger but still subdominant, with a fitted cost of $1.18 \pm 0.01$ ns per launched photon.  This is consistent with the approximately \SI{3.9}{\percent} total detection efficiency and the compact \SI{64}{B} \code{sphoton} hit record: only detected photons, not all transported photons, are returned to the host.

Combining the fitted components gives the empirical wall time model
\begin{equation}
  T_{\mathrm{wall}}(N_{\mathrm{phot}})
  \simeq
  0.31~\mathrm{s}
  +
  \left(147~\mathrm{ns}\right) N_{\mathrm{phot}},
  \label{eq:wall-cost-model}
\end{equation}
where $N_{\mathrm{phot}}$ is the number of launched optical photons. For the benchmark geometry, software versions, and RTX 4090 configuration used here, this model reproduces the measured wall times to better than \SI{5}{\percent} over the tested range. The fitted slopes in table~\ref{tab:timing-fit} show
the corresponding stage-by-stage estimate when the host simulation, GPU optical propagation, or hit return is considered separately.
\section{Application: high throughput optical calorimetry scans}
\label{sec:calorimetry}

The validation and timing results above show that full per-photon optical transport can be used not only for single event checks, but also for multi-parameter detector development studies. As an example, we performed a single particle optical response scan in the benchmark geometry of section~\ref{sec:geometry}. The scan used $e^-$, $\pi^-$, and $p$ primaries
at kinetic energies of 200, 400, 700, 1200, and \SI{3000}{MeV}, with 200 events for each case. The full sample therefore contains 3000 events:
\[
  3~\mathrm{species}
  \times
  5~\mathrm{energies}
  \times
  200~\mathrm{events}.
\]
Each event was simulated with explicit GPU optical photon transport, rather than with a lookup table or parametrized optical response.

The complete scan finished in \SI{3.83}{h} on a single NVIDIA RTX 4090, including the host \Geant{} particle simulation, GPU optical transport, hit return, and output writing.  Using the single-thread \Geant{} optical reference described in section~\ref{sec:performance}, the optical transport component of the same workload would require approximately 10 CPU thread days.
This comparison is intended to illustrate the scale of the acceleration for detector development scans. It is not a production cost estimate for a full experiment simulation chain, which would require a multi-threaded CPU
baseline, experiment specific I/O, and the corresponding reconstruction and
workflow overheads.

The motivation for this type of scan is closely related to light calorimetry
studies in large liquid argon TPCs.  Ning \emph{et al.} recently studied the light side calorimetric response of an idealized LArTPC to GENIE $\nu_e$ charged current events in the \SIrange{0.5}{5}{GeV} range
\cite{ning2025lartpc}. Their study found that the LArTPC light response can be approximately self-compensating, with $e/h \approx 1.0$ at \SI{0.5}{kV/cm}, unlike the higher value obtained from the charge signal. The physical origin is the $dE/dx$ dependence of recombination: heavily ionizing hadronic deposits lose part of their charge signal to recombination light, which can partly compensate the lower charge response of hadronic energy deposits.

The present scan is not a reproduction of that calorimetry model. Instead, it demonstrates the complementary role of explicit optical photon transport. For a specified photon source, detector geometry, wavelength-shifting model, and optical boundary configuration, \Simphony can generate event-by-event detected light distributions quickly enough to support systematic detector studies. Such scans can be used to test optical layouts, build or validate
lookup tables, generate labeled samples for machine learning response models, and study the sensitivity of optical observables to material and geometry choices.

\begin{figure}[h!]
  \centering
\includegraphics[width=0.92\linewidth]{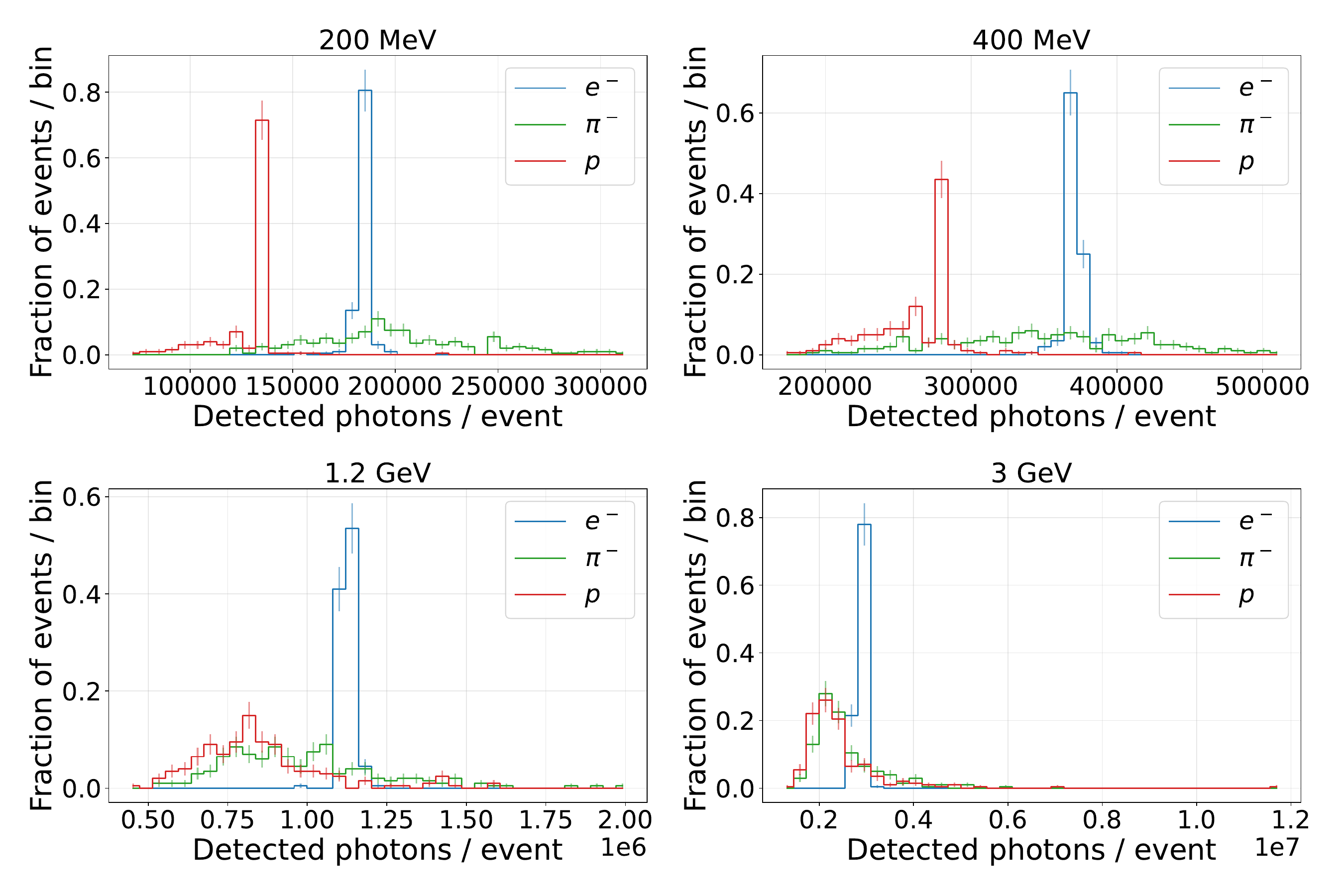}
  \caption{per-event hit count distributions for $e^-$, $\pi^-$, and $p$ primaries in the benchmark geometry. Four representative kinetic energies are shown. Each setting contains 200 events simulated with explicit GPU optical transport. The histograms are normalized separately for each particle species, and the horizontal range in each panel is clipped to the 1st--99th percentile of the pooled distribution for visibility.}
  \label{fig:calorimetry-distributions}
\end{figure}

Figure~\ref{fig:calorimetry-distributions} shows the hit
distributions for the scan at four representative energies. Each histogram is
normalized to the 200 events for the corresponding particle species and
energy. In this simplified benchmark, the lower energy samples show visible species dependent structure in the detected photon count.  The pion samples are broader than the electron samples and contain high yield events, as expected from large fluctuations in hadronic inelastic interactions and, for low-energy $\pi^-$ events, possible nuclear capture final states. At the highest energies shown, the distributions overlap more strongly when only the total detected photon count is used as the response observable.

These distributions should be interpreted as an optical response workflow demonstration, not as a validated detector level calorimetry or
particle identification result.  The benchmark geometry uses idealized wavelength-shifting shells and a 100\%-efficient photon counting boundary, rather than realistic photon detectors. In addition, the validation in this paper is conditional on the generated photon source, it does not validate the absolute scintillation yield, recombination model, quenching model, or hadronic energy response model.  A quantitative calorimetry study would require a validated source term, including an appropriate $dE/dx$-dependent recombination or quenching model such as a Birks, Modified Box, or NEST-like treatment, and comparison with calibration,
test-beam, or detector data.

\section{Scope, limitations, and outlook}
\label{sec:limitations}

The validation in this paper should be interpreted as a controlled optical transport benchmark. It demonstrates that \Simphony reproduces the corresponding \Geant{} optical transport for the simplified large liquid argon geometry, material tables, and WLS model studied here. The comparison is intentionally conditional on the generated optical photon source: the CPU and GPU transports start from the same genstep records, so the test isolates photon propagation, wavelength shifting, optical boundary handling, and hit production. It does not validate the upstream absolute scintillation yield model, recombination model, quenching model, Cerenkov source model, or any detector specific calibration of the absolute light yield.

The benchmark geometry is also intentionally idealized. The active volume is representative of a large LArTPC scale, and the two stage wavelength-shifting shell provides a realistic test for optical transport, but the outer surface is a 100\% efficient photon counting boundary. A detector performance study would need to replace this boundary with realistic photon detection modules, active coverage, mechanical supports, dead regions, SiPM spectral response, quantum efficiency, reflective and absorptive surfaces, wavelength shifter coatings, and measured or constrained surface roughness and reflectivity. The results reported here should therefore not be used to infer an absolute photon detection efficiency, trigger efficiency, or calorimetric performance for a particular experiment.

The timing results have a similarly defined scope. The quoted optical speedups are relative to a single-thread Intel Xeon w7-3445 \Geant{} optical reference. This reference is useful because it yields a clear per thread measure of the optical transport acceleration, but it is not a substitute for a production cost comparison against a fully occupied multi-core CPU node. For experiment scale planning, the next benchmark should include multi-threaded \Geant{} optical transport, realistic event samples, production I/O, experiment-framework overheads, and the scheduling model used on the target computing facility. The present measurements show that the optical transport stage is accelerated by up to about a thousand times in this benchmark.

\section{Conclusions}
\label{sec:conclusions}

We have validated \Simphony for explicit GPU optical photon transport in a simplified \SI{14.7}{kt} liquid argon benchmark containing a two stage wavelength-shifting chain and an idealized photon detector. Starting from common primary photon distributions, the GPU implementation reproduces the corresponding \Geant{} 11.3.2 optical transport for electron shower,
contained muon, and scintillation dominated proton source topologies. The integrated hit ratios remain within \SI{0.25}{\percent} of unity, and the timing, wavelength, and spatial hit comparisons are statistically consistent with the CPU reference.

At the four event \SI{2.5}{GeV} electron operating point, \Simphony transports $243 \pm 3$ million photons in $3.03 \pm 0.06$ s on one RTX 4090. This corresponds to a throughput of $80.2 \pm 1.6$ million photons s$^{-1}$ and an optical transport speedup of $1053 \pm 55$ relative to the single-thread \Geant{} reference. Including upstream event generation, setup, hit return, and output handling, the end-to-end speedup is $89 \pm 5$. The wall time decomposition shows that, after GPU acceleration of the optical transport, the dominant remaining cost in this benchmark is host side event generation.

The 3000 event optical response scan demonstrates that explicit photon transport can be included in detector development parameter scans on a single GPU at the scale tested here. The results should be interpreted as a controlled transport validation rather than a detector performance prediction: realistic photon detector modules, detector specific optical surfaces, calibrated source models, production framework integration, and comparisons with detector data
remain necessary for experiment level applications.

\acknowledgments

This work is supported by the US Department of Energy (DOE) Office of Science, Office of High Energy Physics under Contract No. DE-SC0012704, BNL LDRD 26-055 and BNL LDRD 26-794.

\appendix

\section{One-dimensional projections for muon and proton primaries}
\label{sec:appendix-ratio}

The per-bin Simphony/\Geant ratio analysis of figures~\ref{fig:2p5-ratio-x} for the electron is repeated here for the muon and proton primaries of table~\ref{tab:multipart}. Errorbars are the Poisson 1$\sigma$ uncertainty propagated from both Monte Carlo samples; the dashed line marks unit ratio.

\begin{figure}[h]
  \centering
  \includegraphics[width=0.85\linewidth]{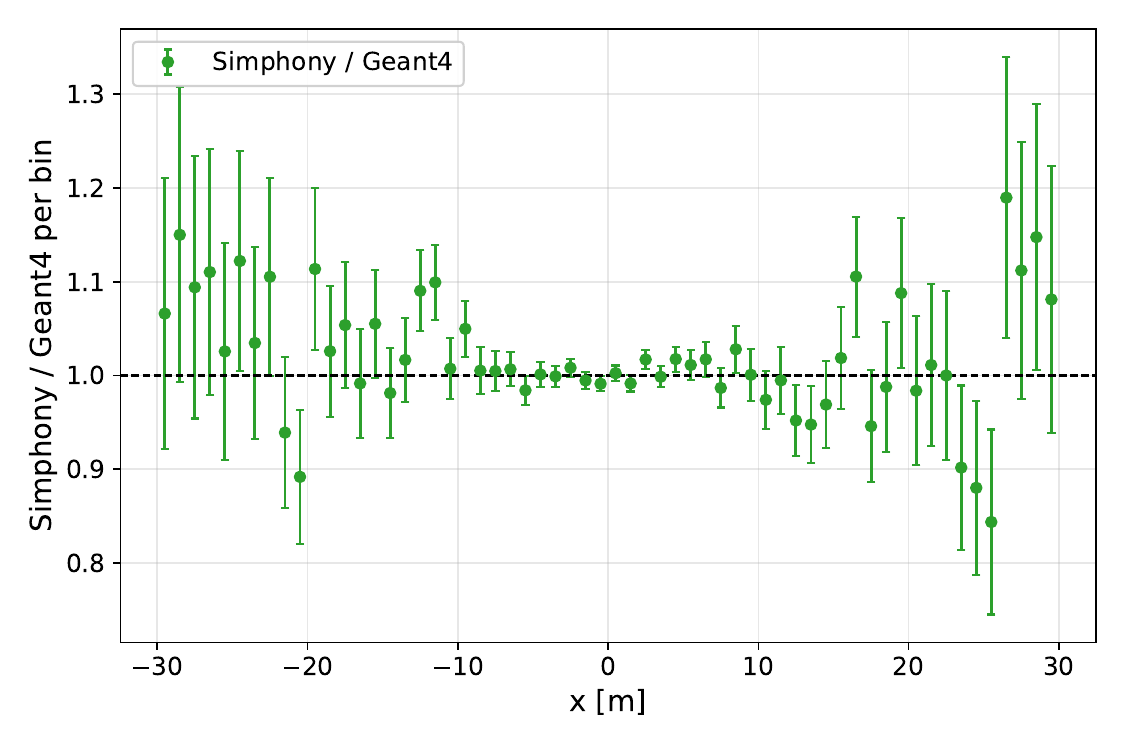}
  \caption{Per-bin ratio along the long $x$ axis, single \SI{1.0}{GeV} muon (seed 42).  $\chisqndf = 0.92$.}
  \label{fig:muon-ratio-x}
\end{figure}

\begin{figure}[h]
  \centering
  \includegraphics[width=0.85\linewidth]{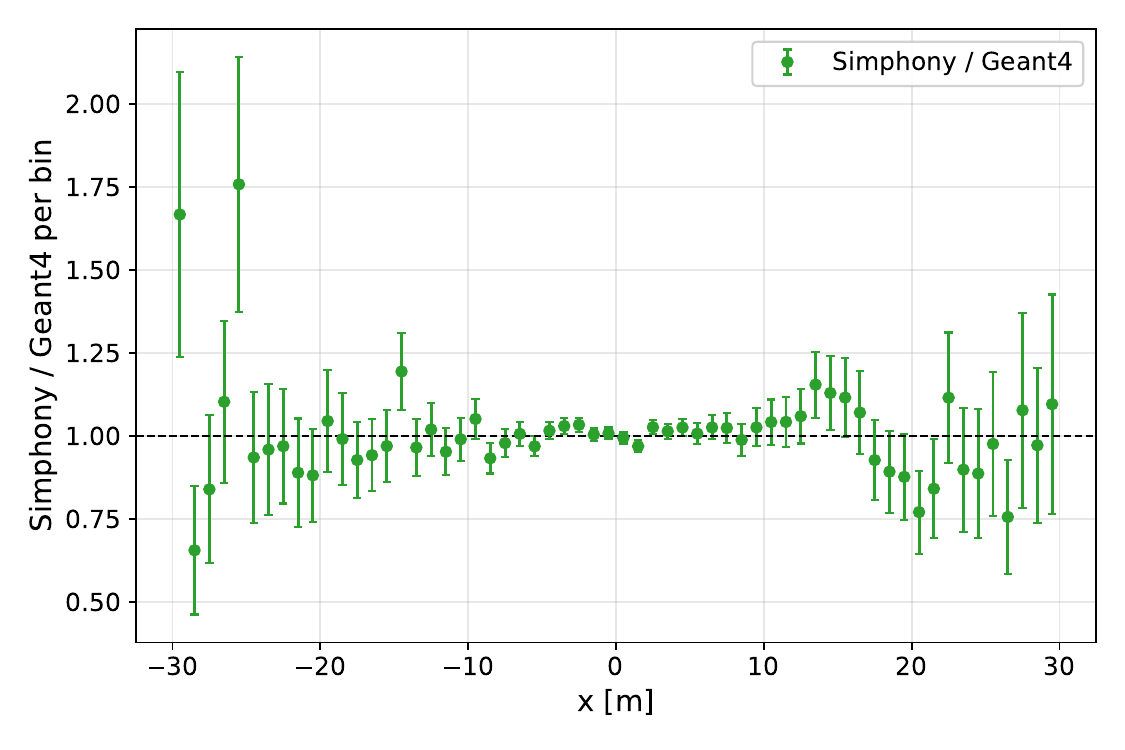}
  \caption{Per-bin ratio along the long $x$ axis, single \SI{400}{MeV} proton (seed 42).  $\chisqndf = 0.80$.}
  \label{fig:proton-ratio-x}
\end{figure}

\clearpage
\section{Optical material properties}
\label{sec:appendix-materials}

The optical material property tables used in the paper benchmark are summarised in
table~\ref{tab:appendix-materials}.  For brevity only the value range and the
energy/wavelength domain over which it is sampled are given. Energies in
the GDML are stored in MeV and lengths in mm following the \Geant{} unit
convention; the table below converts to eV and natural length units.

% Table 9: non-overflowing version
% Requires: booktabs, tabularx, array

\newcolumntype{Y}{>{\raggedright\arraybackslash}X}
\newcolumntype{P}[1]{>{\raggedright\arraybackslash}p{#1}}

\begin{table}[p]
  \centering
  \footnotesize
  \setlength{\tabcolsep}{4pt}
  \renewcommand{\arraystretch}{1.15}

  \caption{Optical material property summary for the benchmark geometry.
  Range notation ``$A\rightarrow B$'' lists the value at the lowest sampled
  energy followed by the value at the highest sampled energy; ``flat'' means
  the property is identical at every sampled point. Length ranges carry
  the domain in eV in parentheses. Energy dependent matrices are sampled on
  the eight point standard grid
  $E\in\{2.34, 2.92, 3.10, 3.65, 4.07, 7.75, 9.69, 11.70\}\,\mathrm{eV}$
  unless otherwise noted.}
  \label{tab:appendix-materials}

  \begin{tabularx}{\linewidth}{@{}P{0.30\linewidth}Y@{}}
    \toprule
    Property & Value or range \\
    \midrule

    \multicolumn{2}{@{}l}{\textbf{Liquid argon} (\texttt{G4\_lAr})} \\
    \addlinespace[2pt]

    RINDEX
      & $1.23\rightarrow 1.6$ over
        $2.34$--$11.7\,\mathrm{eV}$, rising \\

    GROUPVEL
      & $151$--$244\,\mathrm{mm/ns}$, dispersion corrected \\

RAYLEIGH                     & $\approx 0.99\,\mathrm{m}$ at $128\,\mathrm{nm}$ (Babicz \emph{et al.}~2020     
  \cite{babicz2020groupvelocity}); 21-point $\lambda^{4}/(n^{2}-1)^{2}$ dispersion across the LAr active band \\
  ABSLENGTH                    & $80\,\mathrm{m}$ for $E\le 7.5\,\mathrm{eV}$, $10\,\mathrm{m}$ for $E\ge        
  8.0\,\mathrm{eV}$ \\                                                                                             
  SCINTILLATIONCOMPONENT       & truncated Gaussian, peak $128\,\mathrm{nm}$, FWHM $10\,\mathrm{nm}$, window     
  $113$--$155\,\mathrm{nm}$ (Heindl \emph{et al.}~2010 \cite{heindl2010scintillation}); used for                   
  COMPONENT1/COMPONENT2/FAST/SLOW \\

    SCINTILLATIONYIELD
      & $24\,000\,\mathrm{photons/MeV}$ \\

    Time constants, fast/slow
      & $7\,\mathrm{ns}$ / $1400\,\mathrm{ns}$, with yields
        $0.75$ / $0.25$ \\

    Bulk parameters
      & $T=87\,\mathrm{K}$,
        $\rho=1.396\,\mathrm{g/cm^3}$,
        $I_{\mathrm{exc}}=188\,\mathrm{eV}$ \\

    \midrule

    \multicolumn{2}{@{}l}{\textbf{Para-terphenyl} (\texttt{pTP}, inner WLS)} \\
    \addlinespace[2pt]

    RINDEX
      & $1.65$ flat \\

    GROUPVEL
      & $181.7\,\mathrm{mm/ns}$ flat \\

    WLSCOMPONENT
      & Emission peak at $340\,\mathrm{nm}$, with vibronic shoulders at
        $400$--$425\,\mathrm{nm}$ \\

    WLSABSLENGTH
      & $0.5\,\mu\mathrm{m}$ at $128\,\mathrm{nm}$, rising to
        $\ge 10\,\mathrm{m}$ above $340\,\mathrm{nm}$ \\

    ABSLENGTH
      & Shared \texttt{ABSLENGTH\_acrylic} table:
        $1\,\mu\mathrm{m}$ in the VUV $\rightarrow 10\,\mathrm{m}$
        in the visible \\

    WLSTIMECONSTANT
      & $1.136\,\mathrm{ns}$ \\

    \midrule

    \multicolumn{2}{@{}l}{\textbf{TPB-doped acrylic} (\texttt{bluewlsacrylic}, outer WLS)} \\
    \addlinespace[2pt]

    RINDEX
      & $1.58$ flat, EJ-286 PVT base \\

    GROUPVEL
        & $189.7\,\mathrm{mm/ns}$ flat \\

    WLSCOMPONENT
      & Emission peak at $425\,\mathrm{nm}$, with weak
        $530\,\mathrm{nm}$ tail \\

    WLSABSLENGTH
      & $0.1\,\mu\mathrm{m}$ in the VUV;
        $0.8\,\mathrm{mm}$ at $340$--$400\,\mathrm{nm}$ peak;
        $\ge 2\,\mathrm{m}$ above $400\,\mathrm{nm}$ \\

    ABSLENGTH
      & Shared \texttt{ABSLENGTH\_acrylic}, see pTP \\

    WLSTIMECONSTANT
      & $1.26\,\mathrm{ns}$ \\

    \midrule

    \multicolumn{2}{@{}l}{\textbf{SiPM surface}} \\
    \addlinespace[2pt]

    EFFICIENCY
      & $1.0$ flat, idealised photon counter; UNIFIED dielectric--metal,
        polished finish \\

    \bottomrule
  \end{tabularx}
\end{table}

\end{document}